\documentclass[aps,twocolumn]{revtex4}
\usepackage[latin1]{inputenc}
\usepackage{graphicx} 
\usepackage{amsmath}
\usepackage{amsfonts}
\usepackage{amssymb}

\newcommand\bra[1]{\langle\,#1\,\vert}
\newcommand\ket[1]{\vert\,#1\,\rangle}
\newcommand\op[1]{{{\bf #1}}} 
\newcommand\opc[1]{{\cal #1}}

\def \ee {\end{equation}}
\def \be {\begin{equation}}
\def \H {{\cal{H}}}

\def \Hd {{\cal{H}}_D}

\def \dsy {\displaystyle}
\def \barl {\begin{array}{rl}} 
\def \ea {\end{array}}
\newcommand\eq[1]{Eq.~(\ref{#1})}

\let\abs=\envert

\let\norm=\enVert

\newcommand\vers[1]{\op{\hat{#1}}}    % Versor
\newcommand\numeq[1]{(\ref{#1})}
\newcommand\eqs[1]{Eqs.~(\ref{#1})}

\newcommand\sect[1]{Section~\ref{#1}}

\newcommand\bib[1]{\cite{#1}}

\DeclareMathAccent{\mathring}{\mathalpha}{operators}{"17}

\begin{document}

\title{Irreversible decoherence of dipole interacting nuclear spins\\ coupled with a phonon bath} 
\author{F. D. Domínguez,  C. E. González, H. H. Segnorile and R. C. Zamar}
\affiliation {Facultad de Matem\'atica, Astronom\'{\i}a y F\'{\i}sica,
Universidad Nacional de C\'ordoba (FaMAF),  M. Allende y H. de la Torre - Ciudad Universitaria, X5016LAE - C\'ordoba, Argentina. \\
Instituto de F\'isica Enrique Gaviola - CONICET - C\'ordoba, Argentina.}

\begin{abstract}

We report a first-principle theoretical study of the adiabatic decoherence  undergone by  a nuclear spin system in a solid, coupled to the phonon field through the dipolar interaction. The calculations are performed for a chain of weakly interacting 1/2-spin pairs, considered as an open quantum system in contact with a bosonic heat bath. 
By incorporating to the whole system Hamiltonian the fluctuations of the local dipolar energy produced by  low frequency phonons, and assuming that this low energy fluctuations are adiabatic, we find that the spin dynamics can be described in  closed form through a spin-boson model.
The obtained results show that the coupling with the phonons destroy the spin coherence, and the efficiency of the process significantly depends on the complexity of the involved spin states.
By using realistic values for the various parameters of the model, we conclude that this mechanism can be particularly efficient to degrade multi-spin coherences, when the number of `active' spins involved in a given coherence is high. In this way, we show that  the spin coherence in the adiabatic regime can be noticeably affected by this mechanism. 

\end{abstract}
\maketitle

\section{Introduction}      

Quantum dynamics of dipole interacting spin ensembles in solids  arouses great interest in various fields of modern physics, both fundamental and applied.  Particularly, nuclear spins are suitable model systems to face frontier problems on the  physics of many-body systems, like the very foundations of Statistical Mechanics \cite{goldman_talk}. Decoherence and irreversibility are essential pieces for the understanding of the complex dynamics which precedes equilibrium, and they are ultimately linked with basic open questions such as the emergence of thermodynamic equilibrium from the underlying microscopic unitary quantum dynamics \cite{popescu06,deffner2015}. 
Within this context, is the challenging problem of explaining the mechanism that enables spins in solids to attain a quasi-equilibrium state over an early time scale -in its transit to equilibrium- long before the process governing thermalization may have acted. 

In the field of applications, considerable effort is dedicated to manipulating and using quantum spin systems for new applications, like  quantum computations and quantum information processing. A common characteristic of  these developments is that they all need the occurrence of quantum coherence between different states of a many-spin system \cite{Zhang07}. 
Thus, a better understanding of the sources of the environment-induced
destruction of coherent superposition states, that is, of {\em decoherence}, becomes fundamental both in the search of scalable quantum devices and in the characterization of complex (many-body) quantum systems.

A very sensitive technique to probe the nuclear spin dynamics along a wide range of well differentiated timescales is Nuclear Magnetic Resonance (NMR). To our knowledge, no treatment of the irreversible decoherence of solids in NMR has been reported that considers the interacting spins and the environment as a composite system evolving under purely quantum rules. Both basic and applied research fields, would benefit from the development of Hamiltonian models which capture the essential physics and are amenable to rigorous analysis of the many-body nature of the quantum interacting particles coupled to a quantum environment. To advance in this direction it is necessary to adopt a realistic model for the solid, which  allows both to derive a detailed theoretical description from first principles and to elaborate a quantitative analysis of the results. A theoretical approach based on the usual spin-boson model \cite{Legget87,Palma96} that takes account of the spin interactions could meet those expectations. 
Particularly, this strategy could serve to shed light on  the connection between decoherence and system-bath entanglement  of a system of quantum interacting particles \cite{petruccione,Palma96}.

In this work we study the adiabatic decoherence of a system of dipole coupled spins interacting with a phonon bath. The term ``adiabatic'' means that  spins and environment do not exchange energy in the average. 
We follow the basic formal guidelines of the well-known spin-boson model \cite{schloss}, however, instead of considering a set of uncoupled spins interacting individually with the boson field \cite{Fedorov06, priv98}, in our model the observed system is a network of weakly interacting spin pairs,  the boson bath corresponds to the phonons of the lattice at temperature $T$, and the system-environment interaction is generated by the variation of the dipole-dipole energy due to correlated shifts of the spin positions from the equilibrium ones produced by the lattice phonons. For the sake of simplicity of the calculations we keep the main part of this interaction energy coming from the variations of the local (intrapair) energy. 
In other words, we consider an ensemble of spin pairs which are magnetically coupled with all the other pairs and are also correlated with each other through their interaction with the collective modes of the boson bath. In order to facilitate the calculations while keeping the main features of a real system, we considered a unidimensional chain. However, the extension of the analysis to other spin distributions is straighforward. The strategy is based on calculating the exact quantum dynamics of an initial state of the observed system to derive the adiabatic decoherence function of this model. In order to compare with the experiment we estimate the characteristic decoherence time of a typical element of the reduced density matrix.
 
Our calculation aims to inquire if this {\em pair}-boson coupling can act as an effective source of decoherence, able to adiabatically bring an out-of-equilibrium system of weakly coupled pairs to a  state of ``internal'' quasi-equilibrium. 
The occurrence of diagonal states representing quasi-equilibrium states in solid state NMR \cite{Zhang92,Levitt_ernst86,maricq85,Jeene_AdvMR68,eisendrath78,Cho03}) is often postulated on the basis  of the spin temperature hypothesis \cite{abragam61,abragol82,RCZ14}. The spin dynamics associated with these states is described in the framework of `spin thermodynamics', by applying the tools of statistical mechanics to the spins in the quasi-equilibrium state. 

Formally, the statistical properties of the spin system is `kinematically' described by extending to the quasi-equilibrium the mechanical statistical techniques. Thus, a Boltzmann operator involving the various (quasi-) constants of motion or {\em quasi-invariants} is  postulated, as an extension of the equilibrium grandcanonical ensemble technique.
%described with  a Boltzmann operator involving the various (quasi-) constants of motion or {\em quasi-invariants} that can be postulated, as an extension of the equilibrium grandcanonical ensemble technique.
%Formally, the spin thermodynamics of these quasi-equilibrium (diagonal) states is described, within the framework of the grandcanonical ensemble, by a Boltzmann operator which involves the various (quasi-) constants of motion or {\em quasi-invariants}.  
However, the theoretical explanation of the transient processes occurring in the way of the system towards quasi-equilibrium  still represents a challenge for quantum theories of many-body interacting systems. 
Explaining the decay of the off-diagonal density matrix elements could provide additional insight on the idea behind the spin temperature assumption. Similarly, this would allow to shed light on the influence of the environmental degrees of freedom on the generation of multi-spin correlation in the spin dynamics of the solid state \cite{multispBoutis12,Cappellaro_NJP2013,Kroj_Suter04}.

Section \ref{sica} contains a brief review of the main general procedures of the theory of open quantum systems that will be used in the following sections. In Section \ref{modelsys} the spin system is defined, together with the Hamiltonians of the system, bath and spin-bath interaction. Section \ref{dyn} contains a derivation of the decoherence function and an estimation of a characteristic decoherence time scale. It is worth to stress that the simple model system (a chain of dipole coupled spins) used here is not aimed to characterize an actual crystalline sample; it is instead intended to serve as a means to prove that phonons can serve as mediators of an irreversible coherence loss. Finally, the Appendix A is dedicated to validate the adiabatic assumptions used in the calculation of the decoherence function.

\section{Open quantum systems} \label{sica}

As usual when describing an open quantum system in contact with an external bath, we write the total Hamiltonian as
\begin{equation}
\mathcal{H}=\mathcal{H}_S+\mathcal{H}_B+\H_{SL} ,
\label{S-SI-B}\end{equation}
where the system of interest is described (in isolation) by $\mathcal{H}_S$, the bath by $\mathcal{H}_B$ and  $\H_{SL}$ represents the system-bath interaction. 

Since we are interested in describing processes which take place without energy exchange between the spins and the bath, the Hamiltonians must satisfy the {\em adiabatic} condition
\be \label{eqn:adiaba}
\left[ \mathcal{H}_S,\H_{SL}\right]=0 
\ee
which means that the mean value of the spin energy, $\left\langle  \mathcal{H}_S \right\rangle $ is a conserved quantity. 
This condition is essential in this approach since it rules out relaxation and thermalization effects.

To model the interaction Hamiltonian we follow ref. \cite{petruccione2002theory} and write $$\H_{SL}= \sum_{A}\Lambda^A \otimes P^A$$ 
as the sum of products of generic Hermitian operators,  $\Lambda^A$, of the observed system (spins) and  $P^A$ which acts on the Hilbert space of the bath. The only restriction imposed to $\H_{SL}$ is condition (\ref{eqn:adiaba}): $[\H_S,\H_{SL}]=0$. 

The properties of the observed system are expressed by a reduced density operator $\sigma$, which is obtained by tracing the density operator $\rho$ of the whole system over the bath variables \cite{Blum}, that is,  
$$\sigma = \mathrm{Tr}_B \{ \rho \}.$$ 
The strategy for studying decoherence involves calculating the time dependence of $\sigma $.
The unitary dynamics of $\rho$ is driven by the time evolution operator $U(t)= \exp\{-it \H \}$,
which, owing to condition (\ref{eqn:adiaba})  can be factorized as
\begin{equation}\label{eqn:evol}
U(t)=e^{-it\H_S}e^{-it\left( \mathcal{H}_B+\H_{SL}\right) }=V_0(t) \: V(t),
 \end{equation}
where we introduced a ``rigid lattice'' time evolution operator $V_0(t) \equiv e^{-it\H_S}$ which acts on the spin variables only and an operator $V(t)$ which acts on both the spin and lattice variables. Let us further define an operator $V_m(t)$ acting on the {\em lattice} variables only as
$$V(t) \ket{m} =  \ket{m} V_m(t),$$
where $\ket{m}$ is an eigenvector of $\H_{SL}$. 

The time dependence of an element of the reduced density matrix is
\be
\sigma_{mn}(t) =\bra{m}\mathrm{Tr}_B\left\{U(t) \rho(0)\: U^{\dagger(t)} \right\}\ket{n}.
\ee
We now make the usual assumption that the initial condition can be factorized as 
$\rho(0)=\sigma(0)\otimes\rho_B$, then, 
\be
\begin{array}{rl} 
\sigma_{mn}(t) &=\bra{m}V_0(t)\sigma(0)V_0^{\dagger}(t) \ket{n} \: \mathrm{Tr}_B\left\{ V_m(t)\rho_B\: V^{\dagger}_n(t)\right\} 
\end{array} \label{dinamica1} \ee

This very general expression may now be applied to particular cases by selecting the operators $\H_S, \H_{SL}$ and $\H_B$.    
The factor which involves $\sigma(0)$ in Eq.(\ref{dinamica1}) describes the complex (unitary) dynamics of a closed interacting system, while the non-unitary behaviour of this time evolution is given by the trace over the bath variables. Our aim is to calculate the second factor in the case of a linear chain of dipole coupled spin 1/2 pairs interacting with a bath of phonons, as described in Section \ref{modelsys}. 

\section{Model system} \label{modelsys}

\begin{figure}
\includegraphics[width=8cm,height=4cm]{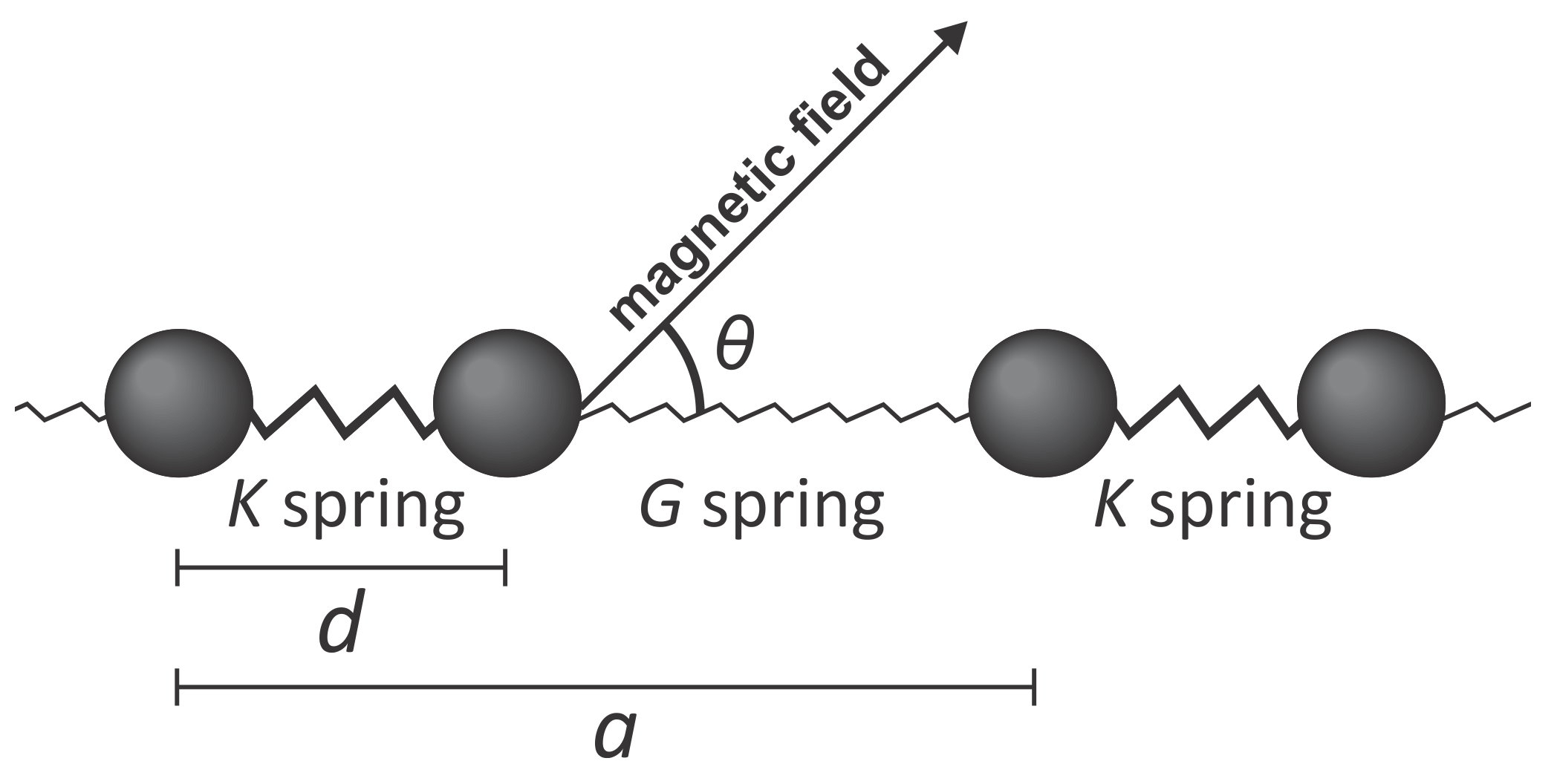}
\caption{This figure represents the model used in the calculations. We assume a monoatomic chain with lattice parameter $a$, with two atoms per unit cell, separated a distance $d$. The chain is embedded in a magnetic field  which form an angle  $\theta$  with the chain.  Each spin interacts with all other spins of the sample through dipolar interaction.}
\label{cadena}
\end{figure}

We consider a system of dipole coupled spin 1/2 pairs in a strong, static magnetic field and represent the model solid as a monoatomic linear chain with a basis. The primitive cell of this unidimensional Bravais lattice with parameter $a$, contains two identical atoms of mass $m$ at a distance $d < a/2$. There are two elastic constants $K$ and $G$ between neighbouring atoms (with $K > G$). See Fig. (\ref{cadena}).

The phonon bath describes the small amplitude displacement of nuclei around their equilibrium positions in the crystal which in turn perturbs their magnetic dipole-dipole interaction. The bath energy of this chain model (in units of $\hbar$) can be written as the sum of uncoupled oscillators of frequency $\omega_k$ (neglecting the zero point energy) as 
\begin{equation} \label{Hbath}
\mathcal{H}_B=\sum _{k}\omega_{k} {b}_{k}^{\dag}{b}_{k},
\end{equation}
where the sum runs over the first Brillouin zone and $b_k$, $b_k^\dag$ are the annihilation and creation operators, respectively, which satisfy $\left[b_k,b_k^\dag\right]=1$.
 The dispersion relation of a 1D chain \cite{Ash} is
\begin{equation}\label{eqn:disp}
{\omega_k}^{2}={\frac {K+G}{m}}\pm \sqrt {{K}^{2}+{G}^{2}+2\,KG\cos \left( ka \right) },
\end{equation}
which admits an optical and an acoustic branch.

In this model each spin interacts with both an external magnetic field (Zeeman energy) and with the other spins of the sample through dipolar interaction.
The Zeeman term is
\begin{equation}
\H_Z=\sum_A\H_Z^A=\sum_{A} \omega_0 \left(I_z^{A1}+I_z^{A2}\right) ,
\end{equation}
where $A$ labels the pair in the chain, and $I_z^{Au}$ is the $z$-component of the angular moment operator of the $u-$th spin in the pair $A$.

As regards the dipolar term, $\Hd$, we first adopt the high magnetic field approximation, since we assume that $\lVert \H_Z \lVert  \gg \lVert \Hd \lVert$ and therefore keep only the secular part of the dipole Hamiltonian, which commutes with  $\H_Z$ (this assumption is very usual in NMR).
Owing to the geometry (chain of pairs) selected in this work, a hierarchy of dipolar couplings arises which naturally allows to split the secular dipolar Hamiltonian into two terms, one involving the interaction between spins belonging to the same cell, which we call {\em intrapair}, and the other representing the {\em interpair} interactions,  $\Hd= \Hd^{intra} + \Hd^{inter}$.  Then we write 
\begin{equation}
\H_{D}^{intra}=\sum _{A=1}^N\H_{D}^A=\sum _{A=1}^N\frac{\Omega_D^A}{3}\left(3 I_Z^{A1}I_Z^{A2}- \boldsymbol{\mathrm{I}}^{A1} \cdot\boldsymbol{\mathrm{I}}^{A2}\right)
\end{equation}
with   $\Omega_D^A$ defined as
\begin{equation}
\Omega_D^A=-\frac{3\mu_0\gamma_P^2\hbar}{8\pi r_A^3}\left[ 3\cos^2\left(\theta \right)-1\right], \label{OmegaDA}
\end{equation}
where  $\theta$ is the angle between the external field and the chain direction, $\gamma_P$ the proton gyromagnetic ratio and $r_A$ is the distance between the interacting nuclei of the $A$ pair.
In the following calculations we will not need the explicit form of the interpair term,  $\Hd^{inter}$, however, we assume that it can be truncated so that it commutes with the intrapair term \cite{keller88}.

We have so far defined the spin interactions and the bath energy. The connection between these two subsystems is provided by the change in the dipolar energy generated by the lattice vibration. It is worth to notice that it will only affect the dipolar and not the Zeeman interaction.
We now assume that this perturbation is small, so that the dipolar energy can be expanded around the equilibrium value of the internuclear distances. In the case of the intra-pair interaction, we have
\be
\mathcal{H}_{D}^A\approx\left.\mathcal{H}_{D}^A\right|_{r_A=d}+\left.\frac{d\mathcal{H}_{D}^A}{dr_A}\right| _{r_A=d}(r_A-d). \label{desarr1}
\ee
The first term of Eq.(\ref{desarr1}) is the {\it rigid lattice} value of the dipolar intrapair Hamiltonian of pair $A$. The second term is the correction due to lattice vibration. Since the aim of this work is to examine if the pair-boson interaction may in fact be a source of decoherence, at this stage, and for the sake of simplicity, we consider only intra-pair variations. This assumption is also based on the fact that $\Hd^{inter}$ is smaller than $\Hd^{intra}$ because of the dependence $r^{-3}$ (see Appendix A for more details). That is,
\be
\mathcal{H}_{D}\approx \left[ \H_{D_0}^{intra}+ \H_{D_0}^{inter}\right] +\sum_A\left.\frac{ d\Hd^{A}}{dr_A}\right| _{r_A=d}(r_A-d),
\label{desarr2}\ee
where $\H_{D_0}$ corresponds to the rigid lattice. 

 We write  $(r_A-d)$ in terms of the displacements from equilibrium, $\mu_{A,u}$, of each spin $u=1,2$ at molecule $A$, since they can be easily related to the phonon creation and annihilation operators $b_{k,s}$ and $ b_{k,s}^\dagger$ of mode $k$ and branch $s$ of an $N$ primitive cells chain, as \cite{Ash}
\begin{equation}
\mu_{A,u}=\frac{1}{\sqrt{N}}\sum_{k,s}\sqrt{\frac{\hslash}{2m\omega_{k,s}}}\left(b_{k,s}+b_{-k,s}^\dagger \right) e^{ik\left( A\cdot a+u\cdot d\right) },
\end{equation}
where $\left( A\cdot a+u\cdot d\right)$ is the equilibrium position of spin $u$ in pair A. Then
\be r_{A}-d=\mu_{A,2}-\mu_{A,1}=\sum_{k}\left( g_k^{A*}b_{k}+ g_k^{A}b_{k}^\dagger\right),
\ee
where we omitted the branch index $s$ to simplify notation, and defined the coupling constants
\be
\label{eqn:consacople}
 g_k^A=e^{-ikAa}g_k,
\ee
with
$$g_k=\left(1-e^{-ikd} \right)\sqrt{\frac{\hslash}{2m\omega_kN}}.$$

In this way the term in square brackets in  Eq.(\ref{desarr2}) is evaluated at the positions of the rigid lattice and the second term, having both spin and lattice variables, emerges as the system-bath interaction. We then define
\be 
\begin{array}{rl}
\H_S &=\mathcal{H}_Z+ \sum _{A}\left.\Hd^A\right|_{d}+\left.\Hd^{inter}\right|_{r_0}, \label{HS_HI} \\ \\
\H_{SL}&=\sum_{A}\Lambda^A\otimes\sum_{k}\left( g_k^{A*}b_{k}+ g_k^{A}b_{k}^\dagger\right)
\end{array}
\ee

where
\be \label{LambdaA}
\Lambda^A=\left.\frac{d\mathcal{H}_{D}^A}{dr}\right| _{r_A=d}=-\frac{\Omega_D}{d}\left(3 I_Z^{A1}I_Z^{A2}- \boldsymbol{\mathrm{I}}^{A1}.\boldsymbol{\mathrm{I}}^{A2}\right)
\end{equation}
is an operator acting on the spin variables only, and
 $\Omega_D=\Omega_D^A|_{r_A=d}$.

Finally, we are going to assume that $[\H_S,\H_{SL}]=0$, wich is equivalent to consider low-frequency modes in the interaction Hamiltonian (see Appendix A for detailed discussion). This allows to factorize the evolution operator as in Eq. (\ref{eqn:evol}).

It is worth to mention that the Hamiltonian $\H_S$ involves the static magnetic dipole interactions between all the spins, both intra-pair and inter-pair. In this way, the time evolution under $\H_S$ only, still reflects the complexity of a multiply connected dipolar network. It is only in $\H_{SL}$ where we neglected the variation of the inter-pair interaction with the displacements. As a consequence, the variations associated to $\Hd^{inter}$, which are not considered in this work, would add an extra source of decoherence.

\section{Dynamics} \label{dyn}
In this section we calculate the exact coherence dynamics given by Eq.(\ref{dinamica1}), using the Hamiltonians of the particular system defined in Eqs.(\ref{Hbath}) and (\ref{HS_HI}).
We also assume that the initial state of the composite system is separable and that the bath is in a thermal state, described by independent density matrices $\Theta_k$ 
\begin{equation}\label{eqn:condini}
\rho(0)=\sigma(0)\otimes\prod_{k}\Theta_k
\end{equation}
with
\be
\begin{array}{rl}
\Theta_k &= Z_k^{-1} e^{-\hbar\beta\omega_kb_k^\dagger b_k} \\ \\
Z_k 	 &=\left( 1-e^{-\beta\omega_k\hbar}\right)^{-1}
\end{array} \ee

In accordance with our definition of   $\H_{SL}$, there exists an eigenbasis $\{\ket{m}\}$ whose elements can be written as direct products of states of the different pairs, 

\begin{equation}
\ket{m}=\ket{m}_1 \otimes \ket{m}_2 \otimes...\ket{m}_N \equiv \ket{m_1,m_2,...,m_N},
\label{state_nonitpairs}
\ee
where each $|m_A\rangle$ is any of the four states of pair $A$, that is, eigenstates of  $\mathcal{H}_{D}^A$ and $\mathcal{H}_Z^A$, 
\begin{equation}
|m_A\rangle\in\left\lbrace |1,1\rangle,|1,0\rangle, |1,-1\rangle,|0,0\rangle\right\rbrace .
\end{equation}
The corresponding eigenvalues of $\Lambda^A$ are
\begin{equation}
\Lambda^A|m_A\rangle=\lambda_{m_A}|m_A\rangle. 
\ee
The action of $\H_{SL}$ over a state $|m\rangle$ is
\be \begin{array}{ll}
\H_{SL}|m\rangle &=\sum_{A}\Lambda^A\otimes\sum_{k}\left( g_k^{A*}b_{k}+ g_k^{A}b_{k}^\dagger\right) |m_1,...,m_N\rangle \\ \\
	&=\sum_{A}\lambda_{m_A}\sum_{k}\left( e^{ikAa}g_k^* b_{k}+ e^{-ikAa}g_k b_{k}^\dagger\right)\ket{m}  \\ \\
	&=\sum_{k}\left( \lambda_{m,k}^*\; g_k^*\; b_{k}+ \lambda_{m,k}\; g_k\; b_{k}^\dagger\right)\ket{m},
\end{array} \label{autovHI} \ee
where we defined 
\be
\lambda_{m,k}=\sum_{A}e^{-ikAa}\lambda_{m_A}. \label{lambda_mk}
\ee
to emphasize that the sum over $A$ in Eq.(\ref{autovHI}) links the eigenvalue of each pair with its position in the chain.%******************
 
We now introduce the notation
\begin{equation} \label{HIm}
\H_{SL}\left( m\right)=\sum_{k}\left( \lambda_{m,k}^{*}g_k^{*}b_{k}+ \lambda_{m,k}g_kb_{k}^\dagger\right),
\end{equation}
which allows writing
\begin{equation}
e^{-it\left( \mathcal{H}_B+\H_{SL}\right) }|m\rangle= e^{-it\left[ \mathcal{H}_B+\H_{SL}\left( m\right) \right]  }|m\rangle\label{evolt_1}
\end{equation}
(remember that $\mathcal{H}_B|m\rangle=|m\rangle\mathcal{H}_B$).
From Eq.\eqref{Hbath} and Eq.\eqref{HIm}, the exponent at Eq.\eqref{evolt_1} can be written as a sum over different modes $k$ as
\be \H_B+\H_{SL}( m) = \sum_{k}\gamma_{k,m}, \ee
where
\begin{equation}
\gamma_{k,m}\equiv \omega_kb_k^{\dagger}b_k+\lambda_{m,k}^{*}g_k^{*}b_{k}+ \lambda_{m,k}g_kb_{k}^\dagger\
\end{equation}
Since terms with different $k$ commute, the time dependence of the density operator of Eq.\eqref{dinamica1} becomes
\be \begin{array}{rl}
 \sigma_{mn}(t)&=\bra{m}V_0(t)\sigma(0)V_0^{\dagger}(t)\ket{n} \\\\
	       & \times \prod_{k}\left\lbrace \mathrm{Tr}_k\left[ e^{-i\gamma_{k,m} t}\Theta_ke^{i\gamma_{k,n} t}\right] \right\rbrace , \label{rhodt}
\end{array} 
\ee
where the trace over the phonon modes $k$ comes from the trace over the states of the heat bath. % in Eq.\eqref{evolt_1}. 
Equation (\ref{rhodt}) has the same structure of Eq.(2.17) from Ref.\cite{priv98}, but here the coefficients $\gamma_{k,m}$ have a definite meaning in terms of the lattice parameters.
The key problem now, is how to calculate such trace. Some authors solve spin-boson problems by calculating an explicit expression for the time evolution operator \cite{Palma96,quiroga02}, an alternative strategy uses the coherent states for the harmonic oscillator to calculate the traces \cite{priv98,Lidar01}. We adopted the latter strategy.

Coherent states $\ket{z} $ are eigenstates of the annihilation operator $b_k$. The trace of an operator in the basis $\{\ket{z}\}$ can be calculated with the following integral \cite{louisell73},
\begin{equation}
\mathrm{Tr}(O)=\int d^2z \bra{z}O\ket{z}, 
\end{equation}
with 
\begin{equation}
d^2z=\frac{1}{\pi}d(\mathrm{Re}\; z)\; d(\mathrm{Im}\; z).
\end{equation}
Using that 
$$
1=\int d^2z\ket{z}\bra{z},
$$
we get for arbitrary $k$
\be \begin{array}{rl}
\mathrm{Tr}_k\left[ e^{-i\gamma_{k,m}t}\right. &\Theta  \left. e^{i\gamma_{k,n}t}\right] = \\ \\
 = & \frac{1}{Z_k}\int d^2z_1d^2z_2d^2z_3 \langle z_1|e^{-i\gamma_{k,m}t}|z_2\rangle \\ \\
   & \times \langle z_2|e^{-\beta \hbar \omega_k b_{k}^{\dagger}b_k}|z_3\rangle\langle z_3|e^{i\gamma_{k,n}t}|z_1\rangle.
\end{array}\label{trace1} \ee

It is well known that \cite{louisell73}
\begin{equation}
\langle z_2|e^{-\beta \hbar \omega_k b_{k}^{\dagger}b_k}|z_3\rangle=e^{z_2^*( e^{-\beta \hbar \omega_k}-1)z_3}.
\end{equation}
In order to calculate the two other matrix elements of the complex exponential operators in Eq.(\ref{trace1}), it is convenient to introduce the Bogoliubov shifted operators  
\begin{equation}
\eta=b+\frac{\lambda_mg}{\omega}
\end{equation}
which allows writing
\begin{equation}
\gamma_m=\omega \eta^{\dagger}\eta-\frac{|\lambda_m|^2|g|^2}{\omega}
\end{equation}
to obtain 
\be \begin{array}{rl}
\langle z_1|e^{-i\gamma_{m}t}|z_2\rangle= & e^{it\frac{|\lambda_m|^2|g|^2}{\omega}}\langle z_1|z_2\rangle \\\\ 
	& e^{\left( e^{-i\omega t}-1\right) (z_1^{*}+\frac{\lambda_m^*g^*}{\omega})(z_2+\frac{\lambda_mg}{\omega})}.
\end{array} \ee

In this way, calculation of the trace in Eq.(\ref{trace1}) implies calculating six gaussian integrals. The result is analogous to the one reported by Ref.\cite{priv98},  that is, 
\begin{equation}
|\sigma_{mn}(t)|=|\bra{m}V_0(t)\sigma(0)V_0^{\dagger}(t)\ket{n}|e^{ -\Gamma_{mn}(t) } \label{dinamica2}
\end{equation}
where $\Gamma_{mn}(t)$ is the {\em decoherence function }

\be \begin{array}{rl}\label{gamma_sum}
\Gamma_{mn}(t)= &2\sum_{k,s}\left|\lambda_{m,k,s}-\lambda_{n,k,s}\right|^2\frac{|g_{k,s}|^2}{\omega_{k,s}^2}\times \\ \\ 
	& \times \left(  1-\cos{\omega_{k,s}t}\right) \coth{\frac{\beta \hbar \omega_{k,s}\hbar}{2}}  
\end{array} \ee 
(we restored the branch index $s$).

\subsection{Analysis of the decoherence function}

In this section we work in the expression of the decoherence function in order to analyze its time dependence and estimate a decoherence time in terms of realistic values of the lattice parameters. 

In the continuum limit we integrate over $k>0$ due to the parity of the integrand, so, Eq.(\ref{gamma_sum}) transforms into
\be \begin{array}{rl}\label{gamma_int} 
\Gamma_{mn}(t)= &\dsy 4\sum_s \int _{0}^{\pi/a}\left|  \lambda_{m,k}-\lambda_{n,k}\right|^2 \frac{|g(\omega_{k,s})|^2}{\omega_{k,s}^2} \times \\ \\ & \dsy \left(  1-\cos{\omega_{k,s}t}\right) \coth{\frac{\beta\omega_{k,s}\hbar}{2}} G(k)dk
\end{array}\ee
where 
\be
G(k)dk=\frac{aN}{\pi}dk
\ee
is the density of states of a linear chain and
\begin{equation}
|g(\omega_{k,s})|^2=\frac{\hbar}{mN\omega_{k,s}}(1-\cos kd).
\end{equation}

\noindent Since the integral in Eq.(\ref{gamma_int}) cannot be calculated analiticaly, we now analize the behaviour of the integrand as a function of $k$. In first place, we consider only the acoustic modes since they are the most populated at room temperature, 
then we use the dispersion relation 
\be
{\omega_k} \sim\sqrt{\frac{KG}{2m(K+G)}}ak \equiv ck 
\ee 
and write the integrand as 
\be 
\begin{array}{rll}
I(k,t) &= &\left|\lambda_{m,k}-\lambda_{n,k}\right|^2 \frac{ 1-\cos(ck t)}{(c k) ^2}   \times  \\ \\
        && \coth\left(\frac{\beta c k\hbar}{2}\right)   \frac{\hbar a}{m \pi}\frac{1-\cos (kd)}{c k} \\ \\
	& \equiv & L_{mn}^2 f(k) h(k,t).
\end{array} \label{integrand}
\ee
We defined the functions $L_{mn}$, $h(k,t)$ and $ f(k) $ to consider their behaviour separately
\begin{itemize}
\item The dependence of $$ L_{mn} \equiv \left|\lambda_{m,k}-\lambda_{n,k}\right| = \left|\sum_{A}e^{-ikAa}(\lambda_{m_A}-\lambda_{n_A}) \right| $$
 on the spin states $\ket{m}$ and $\ket{n}$ is complex. A direct calculation of $L_{mn}$ would involve a sum over the $4^N$ states, then, we must seek for convenient approximations to find an estimation of its possible values. 

The four eigenvalues $\lambda_{m_A}$ of operator $\Lambda^{A}$ (from Eq.\eqref{LambdaA}) of each $A$ pair are
\begin{equation}\begin{array}{rl}
\Lambda^A \ket{1,1} &=0.5\frac{\Omega_D}{d}\ket{1,1} \\\\
\Lambda^A\ket{1,0}  &=-1\frac{\Omega_D}{d}\ket{1,0} \\\\
\Lambda^A\ket{1,-1} &=0.5\frac{\Omega_D}{d}\ket{1,-1} \\\\
\Lambda^A\ket{0,0}  & =0$$
\end{array} \ee

From Eq.\eqref{gamma_int} it is clear that $L_{mm}=0$ and consequently $\Gamma_{mm}=0$, which is a consequence of \eqref{eqn:adiaba}. It is also possible that  $L_{m,n}=0$ for  $m \neq n$  for some particular states, like \mbox{$\ket{m}=|1,1\rangle \otimes |1,1\rangle ... |1,1\rangle$} and  \mbox{$\ket{n}=|1,-1\rangle \otimes |1,-1\rangle ... \ket{1,-1}$}. In a more general case $L_{m,n}$ may involve up to $N$ nonzero terms. 
%In order to proceed with the estimation we have to introduce an approximation for this factor. 
Let us assume for example that only $M$ of the terms contribute to the sum over $A$ and also that the coefficients $(\lambda_{m_A}-\lambda_{n_A})\sim \frac{\Omega_D}{d}$. In such a case
\be \begin{array}{rl}
L_{m,n}^2 & \sim \left(\frac{\Omega_D}{d}\right)^2 \left| \sum_{A=1}^{M}e^{-ikAa}\right|^2 \\ \\
	& = \left(\frac{\Omega_D}{d}\right)^2 \frac{1-\cos\left[  Mka\right] }{1-\cos\left( ka\right) }. \\ \\
\end{array} \label{numberM}
\ee
This expression exposes the fact that $\Gamma_{mn}$ depends on the number $M$ of spins involved in the transition which links $\ket{m}$ and $\ket{n}$ states.

In the limit of small $k$,
$$ \lim_{k\rightarrow 0}L_{mn}^2 = \tilde L_{mn}^2 = \left(\frac{\Omega_D}{d}\right)^2  M^2
$$

\item  $\qquad h(k,t)\equiv \frac{1-\cos(ckt)}{c^2 k^2} = \frac{t^2}{2} {\rm sinc}^2\left(\frac{ckt}{2}\right) $\\ \\
 takes its maximum at $k=0$ and its first minimum (as a function of $k$) at $\dsy k_0= \frac{2\pi}{ct}$. 

\item Concerning $ f(k) \equiv \coth{\frac{\beta\hbar ck}{2}} (1-\cos(kd)) \frac{\hbar a}{m \pi c k},$ \\ \\
 there exists an interval $(0, k_{max})$ where 
\be \begin{array}{rl}
\coth{\frac{\beta\hbar ck}{2}} & \sim \left( \frac{\beta\hbar ck}{2}\right)^{-1}\\ \\  
1-\cos(kd)& \sim\frac{(kd)^2}{2} ,  
\end{array} \label{desarr}
\ee
is valid, and the function $f(k)$ can be approximated by its small-$k$ expression, $\tilde f(k)$. 

\end{itemize}

We now define 
$$\tilde I(k,t)= \tilde L_{mn}\; \tilde f(k) \; h(k,t),$$ which should approximate $I(k,t)$ provided  $k_{max} \gg k_0$ (because $h(k,t)$ becomes very small if $k\gg k_0$). In order to analyze the time range where this approximation is valid, we compare  $I(k,t)$  and $\tilde I(k,t)$ at different times. To do so, it is necessary to set the values of the different constants involved. We choose $d=0.1$nm and $a=3d$, which are the usual values for interatomic distances in solids (for example the distance between Hidrogen nuclei in a hydration water molecule); $m=1.66 \cdot 10^{-27}$kg, is the mass of an Hidrogen atom,  $\Omega_D=300$kHz and $T=300$K and for the speed of sound in a solid we use $c=3000$m/s.

We can now proceed to calculate the exact and the approximate integrands, assuming, for example, $M= 10^3$.  Figure \ref{fig1} shows the plots of  $I(k,t)$ (solid) and $\tilde I(k,t)$ (dashed), calculated at $t=1$ps in Fig.\ref{fig1}(a) and $t=1$ns in Fig.\ref{fig1}(b). It is evident that the two functions coincide in the lower plot, showing that the condition $k_{max} \gg k_0$ is valid for times $t \geq 1 $ns. 
\begin{figure}
\includegraphics[width=8cm,height=11.5cm]{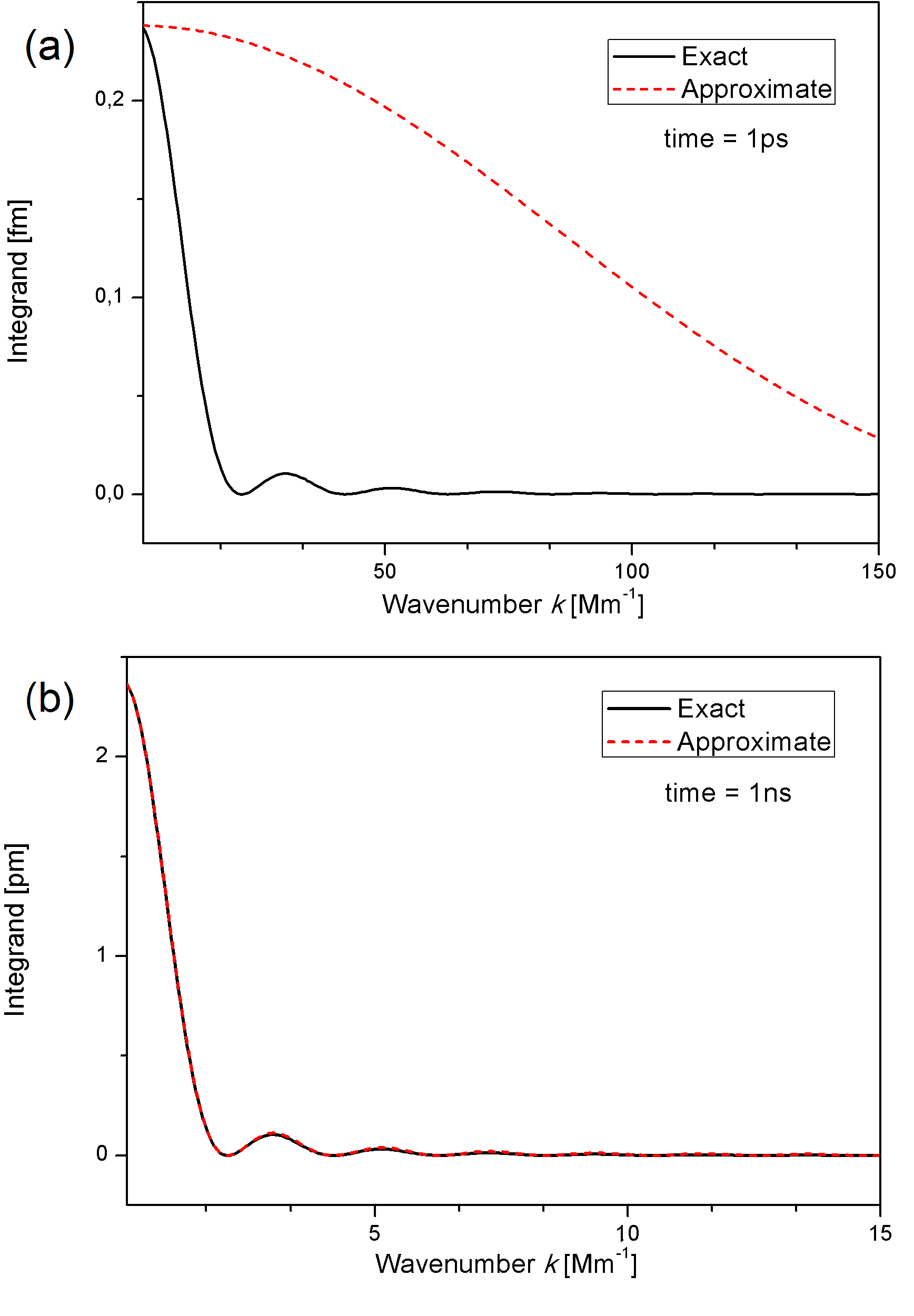}
\caption{(color online) Plot of function $I(k,t)$ from Eq.(\ref{integrand}) (solid line) and the approximate function $\tilde I(k,t)$ (dashed) for two different times: (a) $t= 1\times 10^{-12}$s  and (b) $t= 1\times 10^{-9}$s. Both lines coincide in (b).}
\label{fig1}
\end{figure}

Therefore,  under these conditions it is possible to  extend the upper limit of the integral in Eq.(\ref{gamma_int}) to get an approximate decoherence function 
\be \begin{array}{rl} \label{estimat}
\Gamma     &=\dsy \frac{2\Omega_D^2k_BTaM^2}{\pi mc^4}\int_{0}^{\infty}\frac{1-\cos(ckt)}{k^2}dk \\ \\
	   &=\dsy \frac{\Omega_D^2M^2k_BTa}{mc^3} \: t  \\ \\  
       &\simeq 2.5 \times 10^{-3}M^2\frac{1}{\textrm{s}} \: t .
\end{array} \ee
This numerical result implies that 
coherences in this particular spin system, decay in a characteristic time within the millisecond range if $M=10^3$, which is an experimentally accessible time scale. Besides, a most relevant feature of our result is that the decoherence rate is parameterized by the eigenvalues of the spin part of the interaction Hamiltonian.
 Eq.(\ref{estimat}) contains also an explicit dependence on temperature, however, there can also be an implicit dependence of other parameters for example due to the finite phonon lifetime and the speed of sound in the particular system.

Although we made a realistic approximation, it is important to show that the decoherence function diverges, independently of such approximations and of the chosen values of the parameters.%, which implies that phonons can produce an irreversible coherence loss in a chain of dipole coupled spins. 

\vspace{0.5cm}
\textbf{Lemma}: $\lim_{t\rightarrow\infty}\Gamma_{mn}(t)=\infty \quad$ if $\left|  \lambda_{m,k}-\lambda_{n,k}\right|\neq 0$.

\textit{Proof} Let us first split the contributions from acoustic and optical phonons.
 \begin{equation}
\Gamma=\Gamma_{ac}+\Gamma_{op}%>\Gamma_{ac} cuidado con esta
\end{equation}
Since both terms are positive (because their integrands are positive), it suffices to prove that any of them diverge, then we consider $\Gamma_{ac}$.  In order to find a lower bound for the integral of Eq.(\ref{gamma_int}), we see that for an arbitrary wavelength $\epsilon$, 
\be
\Gamma_{ac} \propto \int_{0}^{\epsilon}(...)dk+\int_{\epsilon}^{\pi/a}(...)dk > \int_{0}^{\epsilon}(...)dk.
\ee
By setting $\epsilon$ small enough, it is possible to replace the integrand by its low $k$ value, $\tilde I(k,t)$
which results in
\be\begin{array}{rl} \label{gamma_ac1} \dsy
\int_{0}^{\epsilon}(...)dk \propto& \dsy \int_{0}^{\epsilon}\frac{1-\cos(ckt)}{k^2}dk \\ \\
	=&\dsy \frac{-1+\cos(\epsilon ct)+\mbox{Si}(\epsilon ct)\; \epsilon ct }{\epsilon}%\rightarrow \infty {\rm if} t\rightarrow\infty
\end{array} \ee
where {\rm Si} stands for the sine integral. Since the second line of Eq.\eqref{gamma_ac1} tends to infinity when $t \rightarrow \infty$, then
\begin{equation}
\Gamma_{mn}(t)>\frac{-1+\cos(\epsilon ct)+\mbox{Si}(\epsilon ct)\; \epsilon ct }{\epsilon}\rightarrow\infty \qquad {\rm if }\; t\rightarrow\infty %_{{\rm if} t\rightarrow\infty}
\end{equation}

\begin{flushright}
$\blacksquare$
\end{flushright}

This interesting result shows that the matrix elements of Eq.\eqref{rhodt} attenuate because the lower bound for $\Gamma(t)$ attains large values as time grows, so proving that decoherence due to the interaction of a chain of dipole-coupled-pairs of spins with a phonon bath is possible. 
%
%\section{botin confuso y dudoso}
%\begin{itemize}
%  
%\item The dynamics of coherence loss of entangled many particle clusters has attracted much attention recently. \cite{FedorovF06}
%\item (Schlosshauer 2007), energy dissipation and decoherence are different phenomena,
%\item in our understanding of non-equilibrium quantum many-body phenomena
%\item In this work, we address the role of bosons as mediators of interactions between particles 
%\end{itemize}

\section{Discussion} \label{discussion}

The main result of this work is the finding that a chain of weakly coupled spin pairs, adiabatically interacting with a phonon bath, treated as an open quantum system, can indeed undergo irreversible decoherence over an intermediate time scale, earlier than that dictated by thermal processes. 
The possibility of completing (attaining) a closed derivation of the decoherence process in a realistic model gives us the opportunity of highlighting the main features of this subtle topic that can be expected to find in a real system. Therefore, this interesting academic example can provide insight into the problem of irreversible spin dynamics in solids.
The proposed model  can be generalized to be applied in a real system of weakly coupled pairs, like hydration water molecules in hydrated salts, and could also be extended to ordinary solids. 
The explanation of irreversible decoherence in solids has remained as an open question for a long time in the field of solid state NMR. Particularly, the mechanism by which nuclear spins of a solid are able to achieve a state of quasi-equilibrium continues to be elusive nowadays. The conclusions obtained in this work can shed some light on this question.

The formal aspect of our work is inspired by the  spin-boson model strategy for treating the time evolution operator in the case where the interaction and bath Hamiltonians do not commute. In this framework it is possible to derive the time dependence of the reduced density matrix in `exact way', without resorting to coarse-graining procedures. 
The traditional spin-boson model considers individual spins  interacting with a boson field, and the spin-environment interaction is represented through a quantum Hamiltonian linear in the boson variables, involving phenomenological coupling strengths, 
which naturally satisfies the adiabaticity condition given the form of the system Hamiltonian. The effect of the unkown coefficients on the result is introduced {\it a posteriori} by modeling the spin-bath interaction. In contrast, in our proposal the system is thought as a network of dipole-dipole interacting spins, whose coupling with the environment originates in the small variations of the local spin interaction produced by lattice phonons. Thus, the system-environment interaction is represented by a first-principle dipole-phonon Hamiltonian, in which the physical magnitudes are expressed in a fundamental way. This characteristic of the model may facilitate a close comparison  with experiments.
 Even when the variations of the inter-pair dipolar interactions were not included in the system-environment Hamiltonian, the mechanical correlation of the different pairs established by the phonon displacement field is reflected in the final result as a collective feature of the decoherence process. A similar  discussion was given in the study of decoherence in nematic liquid crystals , where even when the molecules are magnetically isolated in average and the main contribution is calculated using the intramolecular spin-environment Hamiltonian, the calculated decoherence function reflects the influence on a given molecule exerted by the other molecules through their mechanical interaction as a quantum effect \cite{SegZam11,SegZam13}.

It can be seen from Eqs.(\ref{dinamica1}) and (\ref{dinamica2}) that the decoherence mechanism and the multi-spin dynamics of the observed system are complementary aspects of the irreversible dynamics of a system observable. Because of the multi-spin character of $\H_S$, states $|m \rangle$ and $|n \rangle$ are not eigenstates of the evolution operator $V_0(t)$, then the closed-system dynamics represented by $V_0(t) \sigma(0) V_0 ^\dagger (t)$, connects eigenstates of $H_I$ where the number of ``active'' spins (those which change their state in an $m \rightarrow n$ transition) can be large.  This correlation grows with time \cite{multispBoutis12,ChoCappe06} at a rate which depends on the particular dipolar network. %The decoherence rate of $\sigma_{mn}(t)$ which depends in turn on the number of active spins.

By other hand, the decoherence function $\Gamma_{mn}$ also depends on the number of active spins and is responsible of the irreversible decay of the matrix element  $\sigma_{mn}$. This microscopic irreversibility could convey an irreversible behaviour to an observable $\langle \opc O (t) \rangle =  \mathrm{Tr}\left[ \sigma(t)  \opc O\right] $. 
In NMR experiments the observed signal decay during the evolution under the spin interactions is usually interpreted as a consequence of the transformation of single-quantum coherence terms  to unobservable multiple-spin, single-quantum coherence terms, during the unitary evolution of the initial spin density matrix  in a closed system \cite{baumunowitz85}. In such case, the system evolution would be completely reversible and it would be possible, in principle, to perfectly refocuse the spin dynamics. Due to the same reason, a quasi-equilibrium state would not be possible. However, the system is not completely isolated, and depending on the efficiency of both processes, namely the correlation growth and the decoherence effects, the microscopic information leakage can produce an irreversible macroscopic signal decay.
The estimation of the decoherence time scale was made by using realistic values of the various parameters involved. The order of magnitude obtained in Eq.(\ref{estimat}), showed to be  consistent with decoherence rates measured in NMR refocusing experiments, like the `magic echo' and MREV pulse sequences \cite{multispBoutis12,SegZam13}. Also, the obtained values are compatible with the time scale associated with the generation of quasi-equilibrium \cite{jeener1967nuclear}. 

This suggests that adiabatic decoherence due to the {\it pair}-phonon interaction should also be considerded as a 
source of authentic irreversible attenuation of refocused echoes in NMR experiments. This result contrasts with the null expected influence of dipole-phonon interaction in the processes of spin relaxation and thermalization \cite{abragam61,ChoCappe06,Dolinsek00}. This fact should not be surprising since decoherence and thermalization are processes of very different nature.
Thermalization is generally described by Markovian master equations for the reduced density matrix that evolves in a coarse-grained time scale, with a rate controlled by thermal fluctuations of the spin-lattice interaction. On the contrary, adiabatic decoherence is a full-quantum process which involves, as a prime ingredient, the correlation between system and environment generated by the non-commutation of the interaction and bath Hamiltonians, $[\H_{SL},\H_B] \neq 0$.  

We saw that the low frequency phonons are which play the main role in the decoherence function. Besides, the spin-bath coupling associated with these modes guarantees the existence of an adiabatic time scale where the coherence is lost without appreciable influence of thermal fluctuations. This strong relation between decoherence and low-energy excitations remains to be investigated in solid state NMR.
In summary, we found that the dipole-phonon interaction provides an effective  mechanism of adiabatic decoherence for a system of weakly interacting spin pairs in the solid state, even when the thermalization times can be very long. The model presented allows the estimation of decoherence rates in terms of parameters of the system. The system-environment coupling is introduced through the variations that the phonon field produces on the intra-pair interaction energy, considered as the main contribution. Though the interaction Hamiltonian is a sum of  individual terms, the strong correlation that exists between the pairs due to the phonon field is a decisive feature of the decoherence function (like in the case of liquid crystals). 

The adiabatic decoherence induced by the coupling with the environment 
can be considered the initial irreversible mechanism responsible for the degradation of the signal in the intermediate time scale, long before the thermal fluctuations lead the system towards the final equilibrium with the bath. This mechanism can explain  the signal decay in refocusing experiments and also the occurrence of quasi-equilibrium states over the intermediate time scale. 
 
\section{Acknowledgement}

This work was supported by Secretaría de Ciencia y Técnica, Universidad Nacional de Córdoba, and for MINCyT Córdoba. H. H. S. and F. D. D. thank CONICET for financial support.

\appendix

\section{The adiabatic approach}\label{ap:Adiab_Appr}

This appendix is dedicated to validate the adiabatic approach used in the calculation presented. In such approach, the Hamiltonian of the dynamics can be replaced by an adiabatic one, which rules the dynamics of the system in the decoherence time-scale (i.e. the time-scale far earlier than the thermalization one).

The section is divided in four parts. In \ref{ap:Complete_H}, we present the definition of the complete Hamiltonian of the dynamics and the chosen writing for the parts that compound it. In \ref{ap:Lead_EvolOp}, is described an analytical calculus of the approximations used for the dynamics. In \ref{ap:Decoh_H}, we discuss who is the adiabatic Hamiltonian of the dynamics, which is obtained from the complete Hamiltonian.
%In \ref{ap:Truncated_H}, is discussed the form of the truncated dipolar interpair Hamiltonian and what is its effect over the spin dynamics, in particular we observe the here called `clustering effect'.

\subsection{Description of the complete Hamiltonian of the dynamics}\label{ap:Complete_H}

The complete Hamiltonian of the dynamics is formed by the following Hamiltonians: Zeeman, complete dipolar, the interaction and the bath or environment one, that is

\begin{equation}\label{Hamilt_dyn}
\opc{H} \equiv \opc{H}_Z + \opc{H}_D + \opc{H}_I + \opc{H}_B,
\end{equation}
which will be described hereupon.\\
The Zeeman and the dipolar Hamiltonians applies only over the spin-space states, we can write them as

\begin{equation}\label{Hamilt_Z}
\opc{H}_Z \equiv \opc{H}^{(s)}_Z \otimes \op{1}^{(b)},
\end{equation}
\begin{equation}\label{Hamilt_D}
\opc{H}_D \equiv \sum^{2}_{n = -2} \left(\opc{H}^{n(s)}_{Da} \otimes \op{1}^{(b)}
+ \opc{H}^{n(s)}_{De} \otimes \op{1}^{(b)}\right),
\end{equation}
where
\begin{subequations}\label{Hamilt_Dae}
\begin{equation}\label{Hamilt_Da}
\opc{H}^{n(s)}_{Da} \equiv \sum_A \opc{H}^{n(s)}_{Da(A)},
\end{equation}
\begin{equation}\label{Hamilt_De}
\opc{H}^{n(s)}_{De} \equiv \frac{1}{2}\sum_{A;u = 1,2}\,\sum_{B \neq A;v = 1,2} \opc{H}^{n(s)}_{De(A,u;B,v)}.
\end{equation}
\end{subequations}
Hereafter, the symbols of the form  $\op{O}^{(s)}$ and $\op{O}^{(b)}$ indicate operators acting exclusively on the Hilbert space of the spins (or the system) and the bath (or the environment or lattice), respectively, and $\op{1}$ is the identity operator. In \eqs{Hamilt_D} and \numeq{Hamilt_Dae}, the subscript $Da$ and $De$ represents the $intra$ and the $inter$ parts of the dipolar Hamiltonian, on the other hand, the superscript $n$ indicate its component $0,\pm1\,\text{or}\,\pm2$.
The indexes $A$ and $B$ run over the spin-pair number, $u$ and $v$ run over the number of spin inside each pair (defined by the numbers 1 and 2).\\
The interaction Hamiltonian is derived from the dipolar one and can be expressed as

\begin{equation}\label{Hamilt_I}
\opc{H}_I \equiv \sum^{2}_{n = -2} \left(\opc{H}^{n}_{Ia} + \opc{H}^{n}_{Ie}\right),
\end{equation}
where
\begin{subequations}\label{Hamilt_Iae}
\begin{equation}\label{Hamilt_Ia}
\opc{H}^{n}_{Ia} \equiv \sum_A \opc{H}^{n(s)}_{Ia(A)} \otimes \opc{H}^{(b)}_{Ia(A)},
\end{equation}
\begin{equation}\label{Hamilt_Ie}
\opc{H}^{n}_{Ie} \equiv \frac{1}{2}\sum_{A;u = 1,2}\sum_{B \neq A;v = 1,2} \opc{H}^{n(s)}_{Ie(A,u;B,v)} \otimes \opc{H}^{(b)}_{Ie(A,u;B,v)}.
\end{equation}
\end{subequations}
In the same way that \eq{Hamilt_D}, in \eqs{Hamilt_I} and \numeq{Hamilt_Iae} the subscript $Ia$ and $Ie$ represent the interaction part derived from the $intra$ and the $inter$ parts of the dipolar Hamiltonian, and the superscript $n$ indicate the corresponding component.\\
Finalizing the descriptions of the Hamiltonians, the bath Hamiltonian is writing as

\begin{equation}\label{Hamilt_B}
\opc{H}_B \equiv \op{1}^{(s)}\otimes\opc{H}^{(b)}_B.
\end{equation}

Now, a detailed expression for the Hamiltonians previously shown will be presented.
The well-known Zeeman Hamiltonian is

\begin{equation}\label{Def_Hamilt_Z}
\opc{H}^{(s)}_Z \equiv \omega_0\, \sum_{A} \op{I}^{(s)}_{\op{z}(A)},
\end{equation}
where $\omega_0 = \gamma B_0$ is the Larmor frequency, $\gamma$ is the proton gyromagnetic ratio and $B_0$ is the strength of the static magnetic field, which is applied along the laboratory $\vers{z}$ axis, and $\op{I}^{(s)}_{\op{z}(A)} \equiv \op{I}^{(s)}_{\op{z}(A,1)} + \op{I}^{(s)}_{\op{z}(A,2)}$ is the $\vers{z}$ projection of the total proton spin angular momentum of the $A$-th spin-pair, with $\op{I}^{(s)}_{\op{z}(A,i)}$ as the corresponding angular momentum of the $i$-th spin inside each pair.\\

To obtain the expressions of the different parts that conform the dipolar and interaction Hamiltonians, we will define several functions and operators, which are written  under the international system of units (SI). First, we define the following functions

\begin{subequations}\label{Def_Funct_D}
\begin{equation}\label{Def_Funct_D0}
\Omega_{0}(r) \equiv \frac{\mu_0\gamma^{2}_P\hbar}{8\pi}\,\frac{\left[1-3\cos^2(\theta)\right]}{r^{3}},
\end{equation}
\begin{equation}\label{Def_Funct_D1}
\Omega_{\pm1}(r) \equiv \pm\frac{\mu_0\gamma^{2}_P\hbar}{8\pi}\,\sqrt{\frac{3}{2}}\,\frac{\sin(2\theta)}{r^{3}},
\end{equation}
\begin{equation}\label{Def_Funct_D2}
\Omega_{\pm2}(r) \equiv -\frac{\mu_0\gamma^{2}_P\hbar}{8\pi}\,\sqrt{\frac{3}{2}}\,\frac{\sin^2(\theta)}{r^{3}}.
\end{equation}
\end{subequations}
Then, the following tensors are defining

\begin{subequations}\label{Def_Tensor_D}
\begin{equation}\label{Def_Tensor_D0}
\begin{split}
\op{T}^{(s)}_{\{2,0\}(A,u;B,v)} &\equiv \frac{1}{\sqrt{6}}\big[3\op{I}_{\op{z}(A,u)}\op{I}_{\op{z}(B,v)}\\
&\qquad\qquad-\vec{\op{I}}_{(A,u)}\cdot\vec{\op{I}}_{(B,v)}\big]^{(s)},
\end{split}
\end{equation}
\begin{equation}\label{Def_Tensor_D1}
\begin{split}
\op{T}^{(s)}_{\{2,\pm1\}(A,u;B,v)} &\equiv \mp \frac{1}{2}\big[\op{I}_{\pm(A,u)}\op{I}_{\op{z}(B,v)}\\
&\qquad\qquad\;+\op{I}_{\op{z}(A,u)}\op{I}_{\pm(B,v)}\big]^{(s)},
\end{split}
\end{equation}
\begin{equation}\label{Def_Tensor_D2}
\op{T}^{(s)}_{\{2,\pm2\}(A,u;B,v)} \equiv \frac{1}{2}\big[\op{I}_{\pm(A,u)}\op{I}_{\pm(B,v)}\big]^{(s)},
\end{equation}
\end{subequations}
also, we are using for the tensors that involve spins into a pair the following notation

\begin{equation}\label{Def_Tensor_Da}
\op{T}^{(s)}_{\{2,n\}(A)} \equiv \op{T}^{(s)}_{\{2,n\}(A,1;A,2)}.
\end{equation}

Therefore, using the previous definitions of functions and tensor, we have for the $intra$-pair dipolar Hamiltonian

\begin{equation}\label{Def_Hamilt_Da}
\opc{H}^{n(s)}_{Da(A)} \equiv \sqrt{6}\,\Omega_{n}(d)\,\op{T}^{(s)}_{\{2,n\}(A)}.
\end{equation}
On the other hand, for the $inter$-pair dipolar Hamiltonian we have

\begin{equation}\label{Def_Hamilt_De}
\opc{H}^{n(s)}_{De(A,u;B,v)} \equiv \sqrt{6}\,\Omega_{n}(r_{A,u;B,v})\,\op{T}^{(s)}_{\{2,n\}(A,u;B,v)}.
\end{equation}

In the same way, the parts of the interaction Hamiltonian can be writing as follow.
The $intra$-pair dipolar Hamiltonian have the following expression for the spin and the bath operators

\begin{subequations}\label{Def_Hamilt_Ia}
\begin{equation}\label{Def_Hamilt_Ian}
\opc{H}^{n(s)}_{Ia(A)} \equiv -\frac{3}{d}\,\opc{H}^{n(s)}_{Da(A)} = -\frac{3\sqrt{6}\,\Omega_{n}(d)}{d}\,\op{T}^{(s)}_{\{2,n\}(A)},
\end{equation}
\begin{equation}\label{Def_Hamilt_Iab}
\opc{H}^{(b)}_{Ia(A)} \equiv \sum_{k,l} e^{-ikAa}\,\left[g^{*}_{k,l}\,\op{b}^{(b)}_{k,l} + g_{k,l}\,\op{b}^{\dagger(b)}_{k,l}\right].
\end{equation}
\end{subequations}
Otherwise, for the $inter$-pair dipolar Hamiltonian we have the following expression

\begin{subequations}\label{Def_Hamilt_Ie}
\begin{equation}\label{Def_Hamilt_Ien}
\opc{H}^{n(s)}_{Ie(A,u;B,v)} \equiv -\frac{3}{r_{A,u;B,v}}\,\opc{H}^{n(s)}_{De(A,u;B,v)},
\end{equation}
\begin{equation}\label{Def_Hamilt_Ieb}
\begin{split}
\opc{H}^{(b)}_{Ie(A,u;B,v)} &\equiv \sum_{k,l} \big[g^{*(A,u;B,v)}_{k,l}\,\op{b}^{(b)}_{k,l}\\
&\qquad\qquad + g^{(A,u;B,v)}_{k,l}\,\op{b}^{\dagger(b)}_{k,l}\big],
\end{split}
\end{equation}
\end{subequations}
with
\begin{equation}\label{Def_g_Ie}
\begin{split}
g^{(A,u;B,v)}_{k,l} &\equiv \sqrt{\frac{\hslash}{2m\omega_{k,l}N}}\,e^{i2kd}\\
&\qquad\times \big[e^{-ik(Ba+vd)}-e^{-ik(Aa+ud)}\big].
\end{split}
\end{equation}

Finally, neglecting the zero point energy, for the bath Hamiltonian we have

\begin{equation}\label{Def_Hamilt_B}
\opc{H}^{(b)}_B \equiv \sum _{k,l}\omega_{k,l}\,\op{b}^{\dagger(b)}_{k,l}\,\op{b}^{(b)}_{k,l}.
\end{equation}

\subsection{Leading evolution operator}\label{ap:Lead_EvolOp}

In this section an approximation of the evolution operator will be presented. Such approximation has the feature of being very close to the actual operator, under the condition that will be discussed.\\

First, we define a general Hamiltonian $\opc{H}$ as the sum of two parts. One main part $\opc{H}_{\alpha}$, which has a norm or eigenvalues with similar order of magnitude that $\opc{H}$, and other perturbative one $\opc{H}_{\beta}$, which has a norm or eigenvalues smaller than $\opc{H}_{\alpha}$ and $\opc{H}$. Therefore, we write
\begin{equation}\label{Op_Evol}
\opc{H} \equiv \opc{H}_{\alpha} + \opc{H}_{\beta}.
\end{equation}

Extracting the Hamiltonian $\opc{H}_{\alpha}$ from the complete evolution operator, namely $U(t)$, we can factorized as
\begin{equation}\label{Op_Evol}
U(t) \equiv e^{-i\,t\,\opc{H}} = e^{-i t\,\opc{H}_{\alpha}}\,\vec{e}^{\,\left[-i\int^t_0 d\tau\,\opc{H}_{\beta}(\tau)\right]},
\end{equation}
where $\opc{H}_{\beta}(\tau) \equiv e^{i \tau\,\opc{H}_{\alpha}}\,\opc{H}_{\beta}\,e^{-i \tau\,\opc{H}_{\alpha}}$,
and $\vec{e}$ represents a development of an evolution operator with integrals which have been ordered in time, where the bigger the possible values of the time in the extreme of the integrals, more to the left the operator depending of that time will be. That is, such ordered evolution operator represents the following development
\begin{equation}\label{Op_Evol_Ord}
\begin{split}
&\vec{e}^{\,\left[-i\int^t_0 d\tau\,\opc{H}_{\beta}(\tau)\right]} \equiv \op{1} - i \int^{t}_{0} d\tau_1\,\opc{H}_{\beta}(\tau_1)\\
&\, - \int^{t}_{0} d\tau_1 \int^{\tau_1}_{0} d\tau_2 \,\opc{H}_{\beta}(\tau_1)\,\opc{H}_{\beta}(\tau_2)\\
&\, + i \int^{t}_{0} d\tau_1 \int^{\tau_1}_{0} d\tau_2 \int^{\tau_2}_{0} d\tau_3\,\opc{H}_{\beta}(\tau_1)\,\opc{H}_{\beta}(\tau_2)\,\opc{H}_{\beta}(\tau_3)\\
&\,+ (-i)^n \int^{t}_{0} d\tau_1 \cdots \int^{\tau_{n-1}}_{0} d\tau_n \,\opc{H}_{\beta}(\tau_1) \cdots \opc{H}_{\beta}(\tau_n)\\
&\,+ \cdots.
\end{split}
\end{equation}
Assuming that the Hamiltonian $\opc{H}_{\beta}$ have a part that commutes with $\opc{H}_{\alpha}$, namely $\opc{\hat{H}}_{\beta}$, and other one which does not do it, namely $\opc{\widetilde{H}}_{\beta}$, then we can write
\begin{equation}\label{Hamilt_beta_parts}
\opc{H}_{\beta} \equiv \opc{\hat{H}}_{\beta} + \opc{\widetilde{H}}_{\beta}.
\end{equation}
Using the definition \numeq{Hamilt_beta_parts} in the integrals of \eq{Op_Evol_Ord}, and defining
\begin{equation}\label{Hamilt_beta_nocomm_t}
\opc{\widetilde{H}}_{\beta}(\tau) \equiv e^{i \tau\,\opc{H}_{\alpha}}\,\opc{\widetilde{H}}_{\beta}\,e^{-i \tau\,\opc{H}_{\alpha}},
\end{equation}
we will obtain for the first three integrals the following development
\begin{subequations}\label{Op_Evol_Int}
\begin{equation}\label{Op_Evol_Int_1}
\int^{t}_{0} d\tau_1\,\opc{H}_{\beta}(\tau_1) = \opc{\hat{H}}_{\beta}\,t + \opc{\widetilde{I}}^{\,0}_{\beta}(t),
\end{equation}
\begin{equation}\label{Op_Evol_Int_2}
\begin{split}
&\int^{t}_{0} d\tau_1 \int^{\tau_1}_{0} d\tau_2 \,\opc{H}_{\beta}(\tau_1)\,\opc{H}_{\beta}(\tau_2) = \frac{(\opc{\hat{H}}_{\beta}\,t)^2}{2}\\
&\quad + (\opc{\hat{H}}_{\beta}\,t)\,\opc{\widetilde{I}}^{\,0}_{\beta}(t) + \opc{\widetilde{I}}^{\,1}_{\beta}(t)\,(\opc{\hat{H}}_{\beta}\,t)
- (\opc{\hat{H}}_{\beta}\,t)\,\opc{\widetilde{I}}^{\,1}_{\beta}(t)\\
&\quad + \int^{t}_{0} d\tau_1\,\int^{\tau_1}_{0} d\tau_2\,\opc{\widetilde{H}}_{\beta}(\tau_1)\,\opc{\widetilde{H}}_{\beta}(\tau_2),
\end{split}
\end{equation}
\begin{equation}\label{Op_Evol_Int_3}
\begin{split}
&\int^{t}_{0} d\tau_1 \int^{\tau_1}_{0} d\tau_2 \int^{\tau_2}_{0} d\tau_3\,\opc{H}_{\beta}(\tau_1)\,\opc{H}_{\beta}(\tau_2)\,\opc{H}_{\beta}(\tau_3)\\
&= \frac{(\opc{\hat{H}}_{\beta}\,t)^3}{3!}
+ \frac{(\opc{\hat{H}}_{\beta}\,t)^2}{2}\,\opc{\widetilde{I}}^{\,0}_{\beta}(t)
- (\opc{\hat{H}}_{\beta}\,t)^2\,\opc{\widetilde{I}}^{\,1}_{\beta}(t)\\
&+ \frac{(\opc{\hat{H}}_{\beta}\,t)^2}{2}\,\opc{\widetilde{I}}^{\,2}_{\beta}(t)
+ \opc{\widetilde{I}}^{\,2}_{\beta}(t)\,\frac{(\opc{\hat{H}}_{\beta}\,t)^2}{2}\\
&+ (\opc{\hat{H}}_{\beta}\,t)\,\opc{\widetilde{I}}^{\,1}_{\beta}(t)\,(\opc{\hat{H}}_{\beta}\,t)
- (\opc{\hat{H}}_{\beta}\,t)\,\opc{\widetilde{I}}^{\,2}_{\beta}(t)\,(\opc{\hat{H}}_{\beta}\,t)\\
&+ (\opc{\hat{H}}_{\beta}\,t)\,\int^{t}_{0} d\tau_1\,\int^{\tau_1}_{0} d\tau_2\,\opc{H}_{\beta}(\tau_1)\,\opc{H}_{\beta}(\tau_2)\\
&- (\opc{\hat{H}}_{\beta}\,t) \left[\frac{1}{t}\int^{t}_{0} d\tau_1\,\tau_1\,\opc{\widetilde{H}}_{\beta}(\tau_1)\,\opc{\widetilde{I}}^{\,0}_{\beta}(\tau_1)\right]\\
&+ \frac{1}{t}\int^{t}_{0} d\tau_1\,\tau_1\,\opc{\widetilde{H}}_{\beta}(\tau_1)\,(\opc{\hat{H}}_{\beta}\,t)\,\opc{\widetilde{I}}^{\,0}_{\beta}(\tau_1)\\
&- \int^{t}_{0} d\tau_1\,\opc{\widetilde{H}}_{\beta}(\tau_1)\,(\opc{\hat{H}}_{\beta}\,t)\,\opc{\widetilde{I}}^{\,1}_{\beta}(\tau_1)\\
&+ \left[\int^{t}_{0} d\tau_1\,\opc{\widetilde{H}}_{\beta}(\tau_1)\,\opc{\widetilde{I}}^{\,1}_{\beta}(\tau_1)\right] (\opc{\hat{H}}_{\beta}\,t)\\
&+ \int^{t}_{0} d\tau_1 \int^{\tau_1}_{0} d\tau_2 \int^{\tau_2}_{0} d\tau_3 \, \opc{\widetilde{H}}_{\beta}(\tau_1)\opc{\widetilde{H}}_{\beta}(\tau_2)\opc{\widetilde{H}}_{\beta}(\tau_3),
\end{split}
\end{equation}
\end{subequations}
where is defined
\begin{equation}\label{Integ_Hamilt}
\opc{\widetilde{I}}^{\,n}_{\beta}(t) \equiv \frac{1}{t^n}\int^{t}_{0} d\tau\,\tau^n\,\opc{\widetilde{H}}_{\beta}(\tau).
\end{equation}

The value of the matrix elements of the operators of the integrals in \eq{Op_Evol_Int} can be calculated and thus we have an order of magnitude of such integrals. In this way, we define an eigenbasis $\{\ket{\alpha}\}$ of the operator $\opc{H}_{\alpha}$ where $\opc{H}_{\alpha}\ket{\alpha} = \alpha\ket{\alpha}$. Accordingly, we calculate the value of integrals like the following
\begin{equation}\label{Integ_gen}
\begin{split}
&\bra{\alpha}\left[\int^{t}_{0} d\tau\,f(\tau)\,\opc{\widetilde{H}}_{\beta}(\tau)\right]\ket{\alpha'} = \\
&\qquad\qquad\qquad\bra{\alpha}\opc{\widetilde{H}}_{\beta}\ket{\alpha'}\int^{t}_{0} d\tau\,f(\tau)\,e^{i \omega_{\alpha,\alpha'} \tau},
\end{split}
\end{equation}
with $\omega_{\alpha,\alpha'} = \alpha-\alpha'$.
It is worth to note that $\bra{\alpha}\opc{\widetilde{H}}_{\beta}\ket{\alpha} = 0$, because that $\opc{\widetilde{H}}_{\beta}$ is the non-commutative part of $\opc{H}_{\beta}$ with respect to $\opc{H}_{\alpha}$, therefore always we have that $\alpha \neq \alpha'$ in the integrals.\\
If the exponential function $e^{i \tau \omega_{\alpha,\alpha'}}$ varies faster than $f(\tau)$, we can approximate
\begin{equation}\label{Integ_approx_val}
\begin{split}
&I(t) \equiv \int^{t}_{0} d\tau\,f(\tau)\,e^{i \omega_{\alpha,\alpha'} \tau}
\simeq \lim_{\tau\rightarrow 0} f(\tau)\,\int^{t}_{0} d\tau\,e^{i \omega_{\alpha,\alpha'} \tau}\\
&\qquad = t \lim_{\tau\rightarrow 0} f(\tau)\,e^{i \omega_{\alpha,\alpha'} t/2}\,\text{Sinc}\left(\frac{\omega_{\alpha,\alpha'} t}{2}\right),
\end{split}
\end{equation}
where is defined
\begin{equation}\label{Def_Sinc}
\text{Sinc}\left(\frac{\omega_{\alpha,\alpha'} t}{2}\right) \equiv \frac{\sin(\omega_{\alpha,\alpha'} t/2)}{\omega_{\alpha,\alpha'} t/2},
\end{equation}
with the properties
\begin{subequations}\label{Prop_Sinc}
\begin{equation}\label{Prop_Sinc_1}
\text{Sinc}(0) = 1,
\end{equation}
\begin{equation}\label{Prop_Sinc_2}
\text{Sinc}\left(\frac{\omega_{\alpha,\alpha'} t}{2}\right)\bigg|_{t \gtrsim \omega^{-1}_{\alpha,\alpha'}} \simeq 0.
\end{equation}
\end{subequations}
For the cases of the integrals in \eq{Op_Evol_Int}, we have that $\lim_{\tau\rightarrow 0} f(\tau) = 0$, in particular $f(\tau) = (\tau/t)^n$ ($n=0,1,2,\dots$) for \numeq{Integ_Hamilt}.
In the cases of our concern, we will have that $\omega_{\alpha,\alpha'} \gg 1$, thus
\begin{equation}\label{Val_Lim_f}
t \lim_{\tau\rightarrow 0} f(\tau) = \bigg\{
        \begin{array}{l}
        t,\, n = 0, \\
        0,\, n \neq 0.
        \end{array}
\end{equation}
Using \eqs{Prop_Sinc} and \numeq{Val_Lim_f}, we obtain for the integrals in \eq{Op_Evol_Int} that
\begin{equation}\label{Integ_approx_values}
I(t) \bigg\{\begin{array}{l}
        = e^{i \omega_{\alpha,\alpha'} t/2}\,t\,\text{Sinc}\left(\frac{\omega_{\alpha,\alpha'} t}{2}\right),\, n = 0, \\
        \simeq 0,\, n \neq 0.
        \end{array}
\end{equation}

Therefore, we conclude that it is possible to neglect for all $t$ the values of the integrals that involves powers of $\tau$, with respect to the other one which does not do it. Accordingly, we can very well approximate
\begin{equation}\label{Op_Evol_Ord_appr}
\begin{split}
&\vec{e}^{\,\left[-i\int^t_0 d\tau\,\opc{H}_{\beta}(\tau)\right]} \cong
\op{1} - i \opc{\hat{H}}_{\beta}\,t - \frac{(\opc{\hat{H}}_{\beta}\,t)^2}{2} + i \frac{(\opc{\hat{H}}_{\beta}\,t)^3}{3!}\cdots\\
&+ \left(\op{1} - i \opc{\hat{H}}_{\beta}\,t - \frac{(\opc{\hat{H}}_{\beta}\,t)^2}{2} + \cdots\right)
\int^{t}_{0} d\tau_1\,\opc{\widetilde{H}}_{\beta}(\tau_1)\\
&+ \left(\op{1} - i \opc{\hat{H}}_{\beta}\,t + \cdots\right)
\int^{t}_{0} d\tau_1\,\int^{\tau_1}_{0} d\tau_2\,\opc{\widetilde{H}}_{\beta}(\tau_1)\,\opc{\widetilde{H}}_{\beta}(\tau_2)\\
&+ \int^{t}_{0} d\tau_1 \int^{\tau_1}_{0} d\tau_2 \int^{\tau_2}_{0} d\tau_3 \, \opc{\widetilde{H}}_{\beta}(\tau_1)\opc{\widetilde{H}}_{\beta}(\tau_2)\opc{\widetilde{H}}_{\beta}(\tau_3)+\cdots\\
&= e^{-i t\,\opc{\hat{H}}_{\beta}}\,\vec{e}^{\,\left[-i\int^t_0 d\tau\,\opc{\widetilde{H}}_{\beta}(\tau)\right]},
\end{split}
\end{equation}
and, using \numeq{Op_Evol_Ord_appr}, then
\begin{equation}\label{Op_Evol_appr}
U(t) \cong e^{-i t\,\opc{H}_{\alpha}}\,e^{-i t\,\opc{\hat{H}}_{\beta}}\,\vec{e}^{\,\left[-i\int^t_0 d\tau\,\opc{\widetilde{H}}_{\beta}(\tau)\right]}.
\end{equation}

Due that the norm of $\opc{H}_{\beta}$ is very smaller in comparison with the norm of $\opc{H}_{\alpha}$, a final simplification can be done in the case that concerns us.
In such case, the value of \eq{Integ_approx_values}, for $n = 0$, applied in the calculation of the matrix elements \numeq{Integ_gen}, brings us the result
\begin{equation}\label{Integ_n0_appr}
\begin{split}
&\left|\bra{\alpha}\left[\int^{t}_{0} d\tau\,\opc{\widetilde{H}}_{\beta}(\tau)\right]\ket{\alpha'}\right|\\
&\qquad\quad = 2\left|\frac{\bra{\alpha}\opc{\widetilde{H}}_{\beta}\ket{\alpha'}}{\omega_{\alpha,\alpha'}}\right|
\sin\left(\frac{\omega_{\alpha,\alpha'} t}{2}\right) \ll 1.
\end{split}
\end{equation}
Observing the integrals in \eq{Op_Evol_Ord_appr}, we can see that they introduce factors that scale the values obtained from the powers of $\opc{\hat{H}}_{\beta}\,t$. Because of the small value of \eq{Integ_n0_appr} as well as the small values obtained of integrating several times expressions like \numeq{Integ_n0_appr}, this scaling produce a dynamics very slow.
Therefore, if we are not interested in the dynamics for very long times, we can neglect in \eq{Op_Evol_Ord_appr} all the terms with integrals in front of the other ones.
Finally, the last assumption permit us to obtain the leading evolution operator as
\begin{equation}\label{Op_Evol_lead}
U(t) \cong e^{-i t\,\opc{H}_{\alpha}}\,e^{-i t\,\opc{\hat{H}}_{\beta}}.
\end{equation}

\subsection{Relevant evolution operator for the dynamics in the decoherence time-scale}\label{ap:Decoh_H}

In this section we will obtain the leading evolution operator in the decoherence time-scale.
In order of that, we are using the results of \sect{ap:Lead_EvolOp} to extract the relevant Hamiltonian of such dynamics.
The process, to obtain that relevant Hamiltonian, consists in splitting the Hamiltonian in a strong and a non-commutative weak part (i.e. non-commutative with respect to the strong part), and then extracting the strong part as was developed in \sect{ap:Lead_EvolOp}. In that way, with this procedure we are defining different adiabatic frames of reference for the dynamics.\\

Using the Hamiltonian \numeq{Hamilt_dyn}, the form of the exact evolution operator of the dynamics is
\begin{equation}\label{Op_Evol_dyn}
U(t)\equiv  e^{-i t\,\opc{H}} = e^{-i t \left(\opc{H}_Z + \opc{H}_D + \opc{H}_I + \opc{H}_B\right)}.
\end{equation}
First, we extract from \eq{Op_Evol_dyn} the Zeeman Hamiltonian.
Thus, we have for the definitions in \eq{Op_Evol} that $\opc{H}_{\alpha} \equiv \opc{H}_Z$ and $\opc{H}_{\beta} \equiv \opc{H}_D + \opc{H}_I + \opc{H}_B$.
The commutative part of $\opc{H}_{\beta}$ with respect to the Zeeman Hamiltonian is $\opc{\hat{H}}_{\beta} = \opc{H}^{0}_D + \opc{H}^{0}_I + \opc{H}_B$, where $\opc{H}^{0}_D \equiv \opc{H}^{0}_{Da} + \opc{H}^{0}_{De} \equiv \opc{H}^{0(s)}_{Da}\otimes\op{1}^{(b)} + \opc{H}^{0(s)}_{De}\otimes\op{1}^{(b)}$ and $\opc{H}^{0}_I \equiv \opc{H}^{0}_{Ia} + \opc{H}^{0}_{Ie}$.
On the other hand, the non-commutative part will be
$\opc{\widetilde{H}}_{\beta} = \sum^{2}_{n = -2, n \neq 0} \left(\opc{H}^{n(s)}_{Da}\otimes \op{1}^{(b)} + \opc{H}^{n(s)}_{De} \otimes \op{1}^{(b)} + \opc{H}^{n}_{Ia} + \opc{H}^{n}_{Ie}\right)$.
It is worth that the bath Hamiltonian $\opc{H}_B$ has eigenvalues with absolute values higher that the Zeeman's ones, but that does not affect the approximation due to $\opc{H}_Z$ and $\opc{H}_B$ commute completely between them.\\
Now, it is assumed the relationship in \eq{Integ_n0_appr}, where $\{\ket{\alpha}\}$ is in this case the eigenbasis of $\opc{H}_Z$ with eigenvalue $\alpha$.
Therefore, in this first adiabatic approximation, the evolution operator of the dynamic \numeq{Op_Evol_dyn} can be very well written as
\begin{equation}\label{Op_Evol_lead_Hz}
U(t) \cong e^{-i t\,\opc{H}_Z}\,e^{-i t\,\left(\opc{H}^{0}_{Da} + \opc{H}^{0}_{De} + \opc{H}^{0}_{Ia} + \opc{H}^{0}_{Ie} + \opc{H}_B\right)},
\end{equation}
which is well known as the `high-field approximation'.\\

A second adiabatic frame and approximation can be done extracting the secular dipolar $intra$-pair Hamiltonian from \numeq{Op_Evol_lead_Hz}.
In this way, we now define $\opc{H}_{\alpha} \equiv \opc{H}^{0}_{Da}$ and
$\opc{H}_{\beta} = \opc{H}^{0}_{De} + \opc{H}^{0}_{Ia} + \opc{H}^{0}_{Ie} + \opc{H}_B$.
The commutative part of $\opc{H}_{\beta}$ with respect to $\opc{H}^{0}_{Da}$ will be written as
$\opc{\hat{H}}_{\beta} = \opc{\hat{H}}^{0}_{De} + \opc{H}^{0}_{Ia} + \opc{\hat{H}}^{0}_{Ie} + \opc{H}_B$, where
$\opc{\hat{H}}^{0}_{De}$ and $\opc{\hat{H}}^{0}_{Ie}$ are the corresponding commutative part of the dipolar and interaction $inter$-pair Hamiltonians, i.e. $\opc{H}^{0}_{De}$ and $\opc{H}^{0}_{Ie}$, with respect to the dipolar $intra$-pair one $\opc{H}^{0}_{Da}$, respectively. It is worth to note that the interaction $intra$-pair and the bath Hamiltonians, i.e. $\opc{H}^{0}_{Ia}$ and $\opc{H}_B$, commute completely with respect to $\opc{H}^{0}_{Da}$. Accordingly, for the non-commutative part we define
$\opc{\widetilde{H}}_{\beta} = \opc{\widetilde{H}}^{0}_{De} + \opc{\widetilde{H}}^{0}_{Ie}$.\\
In order to obtain the form of $\opc{\hat{H}}^{0}_{De}$, here we open a parenthesis and develop this task in the following.
We can see from \eqs{Hamilt_D}, \numeq{Hamilt_Dae}, \numeq{Def_Hamilt_Da} and \numeq{Def_Hamilt_De}, that we have to consider the problem of truncating the operator
\begin{equation}\label{Op_Hamilt_De}
\begin{split}
\op{T}^{(s)}_{De} &\equiv \sum_{A;u = 1,2}\,\sum_{B \neq A;v = 1,2}\,\Omega_{0}(r_{A,u;B,v})\\
&\qquad\qquad\times \op{T}^{(s)}_{\{2,0\}(A,u;B,v)},
\end{split}
\end{equation}
with regard to
\begin{equation}\label{Op_Hamilt_Da}
\begin{split}
\op{T}^{(s)}_{Da} \equiv \sum_A \op{T}^{(s)}_{\{2,0\}(A)}.
\end{split}
\end{equation}
To obtain the effects over the dynamics of truncating, we can consider the case where the distance between pairs is enough far to write
\begin{equation}\label{Far_pairs_appr}
\Omega_{0}(r_{A,u;B,v}) \simeq \Omega_{0}(r_{A,u';B,v'}) = \Omega_{0}(r_{A,B}),\,\forall\,u,u',v,v'.
\end{equation}
That approximation as far-away pairs is not the exact problem, but to consider the pairs with the actual distances is a correction to such approximation due that only the very-closed pairs do not accomplish exactly the relationship \numeq{Far_pairs_appr}. Otherwise, this approach permits us to handle analytically the truncating and to see directly the effects over the dynamics, which are very similar to the exact calculation.
Therefore we can write
\begin{equation}\label{Op_Hamilt_De_appr}
\op{T}^{(s)}_{De} = \sum_{A,B \neq A}\,\Omega_{0}(r_{A,B})\,\op{T}^{(s)}_{\{2,0\}(A,B)},
\end{equation}
with
\[\op{T}^{(s)}_{\{2,0\}(A,B)} \equiv \sum_{u = 1,2}\,\sum_{v = 1,2}\,\op{T}^{(s)}_{\{2,0\}(A,u;B,v)}.\]
From the last, we can see that the truncating problem is the same that the presented by Keller\bib{keller88}, where the truncated part is
\begin{equation}\label{Op_T_AB_trunc_K}
\begin{split}
&\hat{\op{T}}^{(s)}_{(A,B)} \equiv \frac{1}{\sqrt{6}}\bigg[2\,\op{T}_{\{1,0\}(A)}\op{T}_{\{1,0\}(B)}\\
&\quad+\frac{1}{2}\big(\op{T}_{\{1,1\}(A)}\op{T}_{\{1,-1\}(B)}+\op{T}_{\{1,-1\}(A)}\op{T}_{\{1,1\}(B)}\big)\\
&\quad+4\big(\op{T}_{\{2,1\}(A)}\op{T}_{\{2,-1\}(B)}+\op{T}_{\{2,-1\}(A)}\op{T}_{\{2,1\}(B)}\big)\bigg]^{(s)},
\end{split}
\end{equation}
where
\begin{subequations}\label{Def_Tensor_Tx1}
\begin{equation}\label{Def_Tensor_T10}
\op{T}^{(s)}_{\{1,0\}(A)} \equiv \op{I}^{(s)}_{\op{z}(A)} = \op{I}^{(s)}_{\op{z}(A,1)} + \op{I}^{(s)}_{\op{z}(A,2)},
\end{equation}
\begin{equation}\label{Def_Tensor_T11}
\op{T}^{(s)}_{\{1,\pm1\}(A)} \equiv \mp\frac{1}{\sqrt{2}}\op{I}^{(s)}_{\pm(A)} \equiv \mp\frac{1}{\sqrt{2}}\left(\op{I}^{(s)}_{\pm(A,1)} + \op{I}^{(s)}_{\pm(A,2)}\right),
\end{equation}
\end{subequations}
and the operators $\op{T}^{(s)}_{\{2,\pm1\}(A)}$ are defined in \eq{Def_Tensor_Da}.\\
To close this parenthesis, using the ideas exposed in the last paragraphs, we now are able to write the expression for $\opc{\hat{H}}^{0}_{De}$.
Using the following definitions
\begin{equation}\label{Op_Hamilt_De_appr_trunc}
\hat{\op{T}}^{(s)}_{De} \equiv \Omega_{0}(a)\,\Sigma\hat{\op{T}}^{(s)}_{De},
\end{equation}
with $\Sigma\hat{\op{T}}^{(s)}_{De} \equiv \sum_{A}\,\hat{\op{T}}^{(s)}_{De(A)}$, and
\begin{equation}\label{Op_TDeA_trunc}
\hat{\op{T}}^{(s)}_{De(A)} \equiv \sum_{B \neq A}\,\left(\frac{a}{r_{A,B}}\right)^3\,\hat{\op{T}}^{(s)}_{(A,B)},
\end{equation}
where $a$ is the shortest distance between pairs, we obtain therefore
\begin{equation}\label{Hamilt_De_appr_trunc}
\opc{\hat{H}}^{0}_{De} = \frac{\sqrt{6}}{2}\,\Omega_{0}(a)\,\Sigma\hat{\op{T}}^{(s)}_{De} \otimes \op{1}^{(b)},
\end{equation}
with the commutation property $\left[\opc{\hat{H}}^{0}_{De},\opc{H}^{0}_{Da}\right] = 0$.\\
It is worth to mention that, in the analysis presented in this work, we are not concerned for the explicit form of $\opc{\hat{H}}^{0}_{Ie}$.\\
As in the first case, we will assume the relationship in \eq{Integ_n0_appr}, where $\{\ket{\alpha}\}$ is in this case the eigenbasis of $\opc{H}^{0}_{Da}$ with eigenvalue $\alpha$.\\
Finalizing this second adiabatic approximation, we conclude that the evolution operator of the dynamics can be written as
\begin{equation}\label{Op_Evol_lead_Hz_H0Da}
U(t) \cong e^{-i t \left(\opc{H}_Z+\opc{H}^{0}_{Da}\right)}\,e^{-i t\,\left(\opc{\hat{H}}^{0}_{De} + \opc{H}^{0}_{Ia} + \opc{\hat{H}}^{0}_{Ie} + \opc{H}_B\right)}.
\end{equation}

Following the extraction of adiabatic frames from the dynamics, we can perform a third approximation, and a new frame definition, extracting from the evolution operator \numeq{Op_Evol_lead_Hz_H0Da} the truncated dipolar $inter$-pair Hamiltonian $\opc{\hat{H}}^{0}_{De}$.
In order of that, we define $\opc{H}_{\alpha} \equiv \opc{\hat{H}}^{0}_{De}$ and
$\opc{H}_{\beta} = \opc{H}^{0}_{Ia} + \opc{\hat{H}}^{0}_{Ie} + \opc{H}_B$.
For this case, the commutative part of $\opc{H}_{\beta}$ with respect to $\opc{\hat{H}}^{0}_{De}$ will be written as
$\opc{\hat{H}}_{\beta} = \opc{\hat{H}}^{0}_{Ia} + \opc{\hat{H}}^{'0}_{Ie} + \opc{H}_B$, where
$\opc{\hat{H}}^{0}_{Ia}$ and $\opc{\hat{H}}^{'0}_{Ie}$ are the commutative parts of the interaction $intra$-pair and the once-truncated interaction $inter$-pair Hamiltonians, i.e. $\opc{H}^{0}_{Ia}$ and $\opc{\hat{H}}^{0}_{Ie}$, with respect to the truncated dipolar $inter$-pair Hamiltonian $\opc{\hat{H}}^{0}_{De}$, respectively. Besides, we have that $\opc{\hat{H}}^{0}_{De}$ commute with $\opc{H}_B$.
Accordingly, for the non-commutative part we define
$\opc{\widetilde{H}}_{\beta} = \opc{\widetilde{H}}^{0}_{Ia} + \opc{\widetilde{H}}^{'0}_{Ie}$.\\
Here we open a new parenthesis to obtain, in the following, the form of $\opc{\widetilde{H}}^{0}_{Ia}$.
From \eqs{Hamilt_Ia} and \numeq{Def_Hamilt_Ia}, we can write the $intra$-pair interaction Hamiltonian $\opc{H}^{0}_{Ia}$ as
\begin{equation}\label{Def_Hamilt_Ia0}
\opc{H}^{0}_{Ia} \equiv -\frac{3\sqrt{6}\,\Omega_{0}(d)}{d}\,\sum_{k} \op{T}^{(s)}_{k} \otimes \op{\bar{b}}^{(b)}_{k},
\end{equation}
where are defined
\begin{subequations}\label{Def_Op_HIa0}
\begin{equation}\label{Def_Op_HIa0_Tk}
\op{T}^{(s)}_{k} \equiv \sum_A \op{T}^{(s)}_{\{2,0\}(A)}\,e^{-ikAa},
\end{equation}
\begin{equation}\label{Def_Op_HIa0_bk}
\op{\bar{b}}^{(b)}_{k} \equiv \sum_{l}\,\left[g^{*}_{k,l}\,\op{b}^{(b)}_{k,l} + g_{k,l}\,\op{b}^{\dagger(b)}_{k,l}\right].
\end{equation}
\end{subequations}
Therefore, the last truncating problem is equivalent to obtain the truncated part of each $\op{T}^{(s)}_{k}$ in \eq{Def_Hamilt_Ia0} with respect to $\opc{\hat{H}}^{0}_{De}$. If it is defined a maximum value of $k$, namely $\bar{k}$, such that $\abs{\bar{k}Aa} \ll 1$ and therefore $e^{-ikAa} \approx 1,\,\forall\,k\leq\bar{k}$, we can verify from \eq{Def_Op_HIa0_Tk} that
$\op{T}^{(s)}_{k\leq\bar{k}} \cong \sum_A \op{T}^{(s)}_{\{2,0\}(A)} = \op{T}^{(s)}_{Da}$ and thus
$\left[\opc{\hat{H}}^{0}_{De},\op{T}^{(s)}_{k\leq\bar{k}}\right] \approx 0$.\\
Finally, to close the last parenthesis, using the last conclusions we can neglect in \eq{Def_Hamilt_Ia0} the terms with $k>\bar{k}$ and then define
\begin{equation}\label{Def_Hamilt_Ia0_trunc}
\opc{\hat{H}}^{0}_{Ia} \equiv -\frac{3\sqrt{6}\,\Omega_{0}(d)}{d}\,\sum_{k\leq\bar{k}} \op{T}^{(s)}_{k} \otimes \op{\bar{b}}^{(b)}_{k}.
\end{equation}
It is worth to mention that, as we will see in the following, we are not concerned for the explicit form of $\opc{\hat{H}}^{'0}_{Ie}$.\\
Again, we will neglect the evolution operator produced by $\opc{\widetilde{H}}_{\beta}$ assuming the relationship in \eq{Integ_n0_appr}, where $\{\ket{\alpha}\}$ is in this case the eigenbasis of $\opc{\hat{H}}^{0}_{De}$ with eigenvalue $\alpha$.\\
Finalizing this third adiabatic approximation, we conclude that the evolution operator of the dynamics can be written as
\begin{equation}\label{Op_Evol_lead_Hz_H0Da_H0Det}
U(t) \cong e^{-i t \left(\opc{H}_Z+\opc{H}^{0}_{Da}+\opc{\hat{H}}^{0}_{De}\right)}\,e^{-i t\,\left(\opc{\hat{H}}^{0}_{Ia} + \opc{\hat{H}}^{'0}_{Ie} + \opc{H}_B\right)}.
\end{equation}

Now we can assume the following relationship between norms, $\norm{\opc{\hat{H}}^{'0}_{Ie}} < \norm{\opc{\hat{H}}^{0}_{Ia}}$, thus the decoherence produced by the Hamiltonian $\opc{\hat{H}}^{0}_{Ia}$ will be faster than the produced by $\opc{\hat{H}}^{'0}_{Ie}$.
Accordingly, for a shake of simplicity in the analytical calculation of the dynamics, we will neglect the dynamics generated by $\opc{\hat{H}}^{'0}_{Ie}$.\\

Therefore, to conclude this section, we define the relevant evolution operator, which governs the dynamics in the decoherence time-scale, as the adiabatic expression

\begin{equation}\label{Op_Evol_lead_adiab}
U(t) \simeq e^{-i t \left(\opc{H}_Z+\opc{H}^{0}_{Da}+\opc{\hat{H}}^{0}_{De}\right)}\,e^{-i t\,\left(\opc{\hat{H}}^{0}_{Ia} + \opc{H}_B\right)},
\end{equation}
where $\opc{H}_Z$, $\opc{H}^{0}_{Da}$ and $\opc{\hat{H}}^{0}_{De}$ commute between them and with $\opc{\hat{H}}^{0}_{Ia}$ and $\opc{H}_B$, but
$\left[\opc{\hat{H}}^{0}_{Ia}, \opc{H}_B\right] \neq 0$.

%%%%%%%%%%%%%%%%%%%%%%%%%%%%%%%%%%%%%%%%%%%%%%%%%%%%%%%%%%%%%%%%%%%%%%%%%%%%%%%%%%%%%%%%%%%%%%%%%%
\newpage
\bibliographystyle{unsrt}
\bibliography{ref_cr}

\begin{thebibliography}{10}

\bibitem{goldman_talk}
M.~Goldman.
\newblock Thermodynamics games with nuclear magnets.
\newblock {\em Fifth Descartes Lecture presented to the Science Division of the
  Royal Netherlands Academy of Arts and Sciences Amsterdam, The Netherlands},
  October 30, 1995.

\bibitem{popescu06}
A.~Winter S.~Popescu, A.J.~Short.
\newblock Entanglement and the foundations of statistical mechanics.
\newblock {\em Nature Physics}, 2:754, 2006.

\bibitem{deffner2015}
S.~Deffner.
\newblock From spooky foundations.
\newblock {\em Nature Physics}, 11:383, 2015.

\bibitem{Zhang07}
K.A. Al-Hassanieh W.~Zhang, N.~Konstantinidis and V.V. Dobrovitski.
\newblock {\em J. Phys.: Condens. Matter}, 19:083202, 2007.

\bibitem{Legget87}
A.~J. Leggett, S.~Chakravarty, A.~T. Dorsey, Matthew P.~A. Fisher, Anupam Garg,
  and W~Zwerger.
\newblock Dynamics of the dissipative two-state system.
\newblock {\em Rev. Mod. Phys.}, 59:1--85, Jan 1987.

\bibitem{Palma96}
K.~A.~Suominen M.~Palma and A.~Ekert.
\newblock {\em Proc. R. Soc. London}, 452:567, 1996.

\bibitem{petruccione}
Francesco Petruccione and Heinz-Peter Breuer.
\newblock {\em The theory of open quantum systems}.
\newblock Oxford Univ. Press, 2002.

\bibitem{schloss}
M.~Schlosshauer.
\newblock {\em Decoherence and the quantum-toclassical transition}.
\newblock Springer-Verlag, Berlin Heidelberg, 2007.

\bibitem{Fedorov06}
A.~Fedorov and L.~Fedichkin.
\newblock Collective decoherence of nuclear spin clusters.
\newblock {\em J. Phys.: Condens. Matter}, 18:3217/3228, 2006.

\bibitem{priv98}
D.~Mozyrski and V.~Privman.
\newblock {\em J. Stat. Phys.}, 91:787, 1988.

\bibitem{Zhang92}
B.~Meier S.~Zhang and R.R. Ernst.
\newblock {\em Phys. Rev. Lett.}, 69:2149, 1992.

\bibitem{Levitt_ernst86}
D.~Suter M.~H.~Levitt and R.~R. Ernst.
\newblock {\em J. Chem. Phys}, 84:4243, 1992.

\bibitem{maricq85}
M.~M. Maricq.
\newblock {\em Phys. Rev. B}, 31:127, 1985.

\bibitem{Jeene_AdvMR68}
J.~Jeener.
\newblock Thermodynamics of spin systems in solids.
\newblock In John~S. Waugh, editor, {\em Advances in Magnetic Resonance},
  volume~3 of {\em Advances in Magnetic and Optical Resonance}, pages 205 --
  310. Academic Press, 1968.

\bibitem{eisendrath78}
W.~Stone H.~Eisendrath and J.~Jeener.
\newblock Nmr of protons in gypsum. i. experimental proof of the existence of
  four thermodynamic invariants.
\newblock {\em Phys. Rev. B}, 17:47, 1978.

\bibitem{Cho03}
D.~Cory H.~Cho and C.~Ramanathan.
\newblock {\em J. Chem. Phys}, 118:3686, 2003.

\bibitem{abragam61}
A.~Abragam.
\newblock {\em {Principles of Nuclear Magnetism (International Series of
  Monographs on Physics)}}.
\newblock Oxford University Press, 1983.

\bibitem{abragol82}
A.~Abragam and M.~Goldman.
\newblock {\em Nuclear magnetism: order and disorder}.
\newblock International series of monographs on physics. Clarendon Press, 1982.

\bibitem{RCZ14}
R.C. Zamar.
\newblock Quasi-equilibrium states of nuclear spins in nematic liquid crystals.
\newblock {\em J. Spectrosc. Dyn}, in press, 2014.

\bibitem{multispBoutis12}
Steven~W. Morgan, Vadim Oganesyan, and Gregory~S. Boutis.
\newblock Multispin correlations and pseudothermalization of the transient
  density matrix in solid-state nmr: Free induction decay and magic echo.
\newblock {\em Phys. Rev. B}, 86:214410, Dec 2012.

\bibitem{Cappellaro_NJP2013}
A.~Ajoy G.~Kaur and P.~Cappellaro.
\newblock Decay of spin coherences in one-dimensional spin systems.
\newblock {\em New Journal of Physics}, 15:093035, 2013.

\bibitem{Kroj_Suter04}
H.~G. Krojanski and D.~Suter.
\newblock {\em Phys. Rev. Lett.}, 93:090501, 2004.

\bibitem{petruccione2002theory}
Francesco Petruccione and Heinz-Peter Breuer.
\newblock {\em The theory of open quantum systems}.
\newblock Oxford university press, 2002.

\bibitem{Blum}
Karl Blum.
\newblock {\em Density Matrix Theory and Applications}.
\newblock Springer Series on Atomic, Optical, and Plasma Physics, 2012.

\bibitem{Ash}
N.W. Ashcroft and N.D. Mermin.
\newblock {\em {Solid State Physics}}.
\newblock Saunders College, Philadelphia, 1976.

\bibitem{keller88}
A.~Keller.
\newblock Spin-1 behavior of systems of dipolar coupled pairs of spin-½ nuclei.
\newblock volume~12 of {\em Advances in Magnetic and Optical Resonance}, pages
  183 -- 246. Academic Press, 1988.

\bibitem{quiroga02}
D.~Mozyrski and V.~Privman.
\newblock Decoherence of quantum registers.
\newblock {\em Phys. Rev. A}, 65:032326, 2002.

\bibitem{Lidar01}
Z.Bihary D.A.~Lidar and K.B Whaley.
\newblock From completely positive maps to the quantum markovian semigroup
  master equation.
\newblock {\em Chemical Physics}, 268:35, 2001.

\bibitem{louisell73}
William~Henry Louisell.
\newblock {\em Quantum statistical properties of radiation}.
\newblock John Wiley Sons Canada, 1973.

\bibitem{SegZam11}
H.H. Segnorile and R.C. Zamar.
\newblock Quantum decoherence and quasi-equilibrium in open quantum systems
  with few degrees of freedom: Application to 1h nmr of nematic liquid
  crystals.
\newblock {\em The Journal of Chemical Physics}, 135:244509, 2011.

\bibitem{SegZam13}
H.H. Segnorile and R.C. Zamar.
\newblock Quantum irreversible decoherence behaviour in open quantum systems
  with few degrees of freedom: Application to (1)h nmr reversion experiments in
  nematic liquid crystals.
\newblock {\em J. Chem. Phys.}, 139(15):154901, 2013.

\bibitem{ChoCappe06}
H.~Cho, P.~Cappellaro, D.G. Cory, and Ch. Ramanathan.
\newblock Decay of highly correlated spin states in a dipolar-coupled solid:
  Nmr study of caf2.
\newblock {\em Phys. Rev. B}, 74:224434, 7 2006.

\bibitem{baumunowitz85}
J.~Baum, M.~Munowitz, A.~N. Garroway, and A.~Pines.
\newblock Multiple-quantum dynamics in solid state nmr.
\newblock {\em The Journal of Chemical Physics}, 83(5):2015--2025, 1985.

\bibitem{jeener1967nuclear}
Jean Jeener and Paul Broekaert.
\newblock Nuclear magnetic resonance in solids: thermodynamic effects of a pair
  of rf pulses.
\newblock {\em Physical Review}, 157(2):232, 1967.

\bibitem{Dolinsek00}
J~Dolinsek, P.M Cereghetti, and R~Kind.
\newblock Phonon-assisted spin diffusion in solids.
\newblock {\em Journal of Magnetic Resonance}, 146(2):335 -- 344, 2000.

\end{thebibliography}

%\begin{thebibliography}{aeiou}

%\end{thebibliography}
\end{document}